\documentclass[11pt,a4paper]{article}
\usepackage{jheppub}
\usepackage{amsmath,amsfonts,amsthm,graphicx,ulem}
\usepackage{float}

\newcommand{\comment}[1]{}
\newcommand{\bal}{\begin{align}}
\newcommand{\eal}{\end{align}}
\newcommand{\beq}{\begin{equation}}
\newcommand{\eeq}{\end{equation}}
\newcommand\beqa{\begin{eqnarray}}
\newcommand\eeqa{\end{eqnarray}}
\newcommand\bea{\begin{array}}
\newcommand\eea{\end{array}}

\renewcommand{\leq}{\leqslant}

\renewcommand{\ge}{\geqslant}

    \newcommand{\COMMENT}[1]{}
    
    \newcommand{\neqa}{\nonumber\end{eqnarray}}

    \newcommand{\dbar}{{\textstyle \delta}
\lower.03ex\hbox{\kern-0.38em$^{\scriptstyle-}$}\kern-0.05em{}}

\def\a{{\alpha}}

\def\[{\left[}
\def\]{\right]}

\def\s{\sigma}
\def\a{\alpha}

\def\<{\langle}
\def\>{\rangle}

\def\tr{\text{tr}~}

\def\i2{\frac{i}{2}}

\title{ \center{ Three-point correlator of twist-2 light-ray operators in $\mathcal{N}=4$ SYM in BFKL approximation}}
\author[a,b]{Ian Balitsky}
\author[c,d,1]{Vladimir Kazakov%
\note{member of Institut Universitaire de France}}
\author[e]{Evgeny Sobko}

\affiliation[a]{Physics Dept., Old Dominion University, Norfolk VA 23529}
\affiliation[b]{Theory Group, JLAB, 12000 Jefferson Ave, Newport News, VA 23606}
\affiliation[c]{Ecole Normale Superieure, LPT, 24 rue Lhomond,  75231 Paris CEDEX-5,
  France}
\affiliation[d]{Universit\'e Pierre et Marie Curie, Paris-VI, France}
\affiliation[e]{DESY Hamburg, Theory Group, Notkestra{\ss}e 85, 22607 Hamburg, Germany}

\emailAdd{balitsky AT jlab.org}
\emailAdd{evgenysobko AT gmail.com}
\emailAdd{kazakov AT lpt.ens.fr}

\abstract{We present  calculation of the correlation function of three twist-2
operators in the  BFKL limit. The calculation is performed in $\mathcal{N} = 4$ SYM but the result is valid in
other gauge theories such as QCD. The obtained
leading order structure constant is exact for any number of colors.}

\subheader{DESY 15-211}

\usepackage{varioref}
\usepackage{makeidx}
\makeindex
\usepackage{amssymb}

\begin{document}

  \maketitle

\section{Introduction}
The superconformal       \(\mathcal{N}=4\) SYM   theory, together with some of its deformations, represents a unique opportunity to perform a non-perturbative study, at least in the planar 't~Hooft limit, of the OPE properties of conformal theories at four space-time dimensions. This appears to be possible due to the   quantum integrability \cite{Beisert:2010jr}. The spectrum of anomalous dimensions of local operators is considered to be a solved problem, especially with the discovery of its formulation in terms of the Quantum Spectral Curve (QSC) \cite{Gromov:2013pga,Gromov:2014caa}.  However, to complete the picture of OPE in this theory one  needs to learn how to calculate the structure constants. Both anomalous dimensions and structure constants are in general highly nontrivial functions of the coupling constant. While  the former  are now exactly and efficiently computable  at  large \(N_c\), the calculation of the OPE structure constants, in spite of a considerable progress in the last time \cite{Basso:2015zoa,Sobko:2013ema,Gromov:2012uv,Jiang:2015lda,Basso:2015eqa,Eden:2015ija,Costa:2012cb,Bajnok:2015hla}, is still far from  its final goal -- to establish a concise and comprehensive set of non-linear functional equations handy for the numerical and analytic study, in the spirit of  QSC equations.
In this circumstances, similarly to early stages of the study of spectral problem, it is very useful to find examples of explicit calculation of certain structure constants, especially in the approximations which go beyond the perturbation theory or the strong coupling limit \cite{Janik:2011bd,Kazama:2014sxa}. %
\begin{figure}
  \centering
  \includegraphics[scale=0.335]{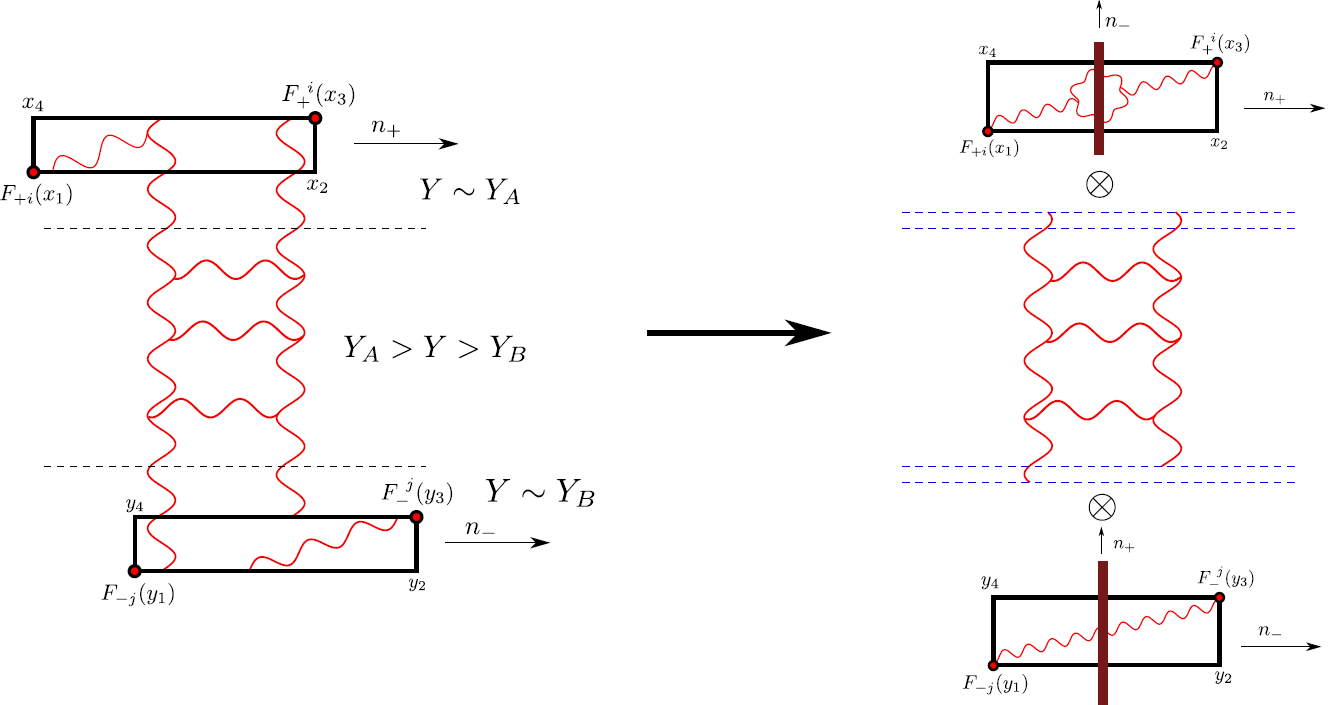}
  \caption{Scheme of computation of 2-point  correlator. In the l.h.s., the long sides of regularizing rectangular Wilson frames are stretched along light ray and the short sides in the orthogonal directions.  In the r.h.s. we use OPE of frames over color dipoles and compute their correlator, see \cite{Balitsky:2013npa} for details.}
  \label{fig:2p}
\end{figure}

In this respect, an interesting opportunity is provided  by the BFKL limit \cite{Fadin:1975cb,Balitsky:1978ic}, widely used in the study of high energy hadron physics. The most direct way to compute the OPE structure constants is to compute the correlation function of the corresponding three operators. In this paper, following our note \cite{Balitsky:2015tca}, we show in detail how to calculate   the 3-point correlator of twist-2 operators \begin{equation} \mathcal{O}^{j}(x)=\text{tr} F_{+i}D_+^{j-2}F_+^i+fermions+scalars\label{twist2op}\end{equation} in \(\mathcal{N}=4\) SYM in the BFKL limit \cite{Fadin:1975cb,Balitsky:1978ic} when the Lorentz spin \(j\) of this operator tends to \(1\): \(\omega=j-1\rightarrow 0\), the 't~Hooft coupling \(g^2\equiv \frac{N_c g_{YM}^2}{16 \pi^2}\to 0\) and the ratio \(\frac{g^2}{\omega}\)~fixed, for arbitrary \(N_c\).
 Since the contribution of {\it\ fermions+scalars} is subleading at this limit (including the internal loops) the result is valid for the pure Yang-Mills theory as well.
  The case of two-point correlator  was elaborated in our previous paper \cite{Balitsky:2013npa} where we  defined  the generalized operators with complex spin as special light-ray operators \cite{Balitsky:1987bk}, regularized by a narrow rectangular Wilson contour stretched in a light-cone direction called "frame". We calculated their two-point correlator using OPE over Wilson lines \cite{Balitsky:1995ub} with a rapidity cutoff and the BFKL evolution (see Fig.\ref{fig:2p}). Here we  use the same light-ray operators: one along \(n_+\) direction and    two, instead of one,  along \(n_-\). It imposes on us the use of a  more general Balitsky-Kovchegov (BK) evolution equation \cite{Balitsky:1997mk,Kovchegov:1999ua,Kovchegov:1999yj} and the leading  BFKL contribution comes from the triple-pomeron vertex \cite{Bartels:1994jj,Korchemsky:1997fy}.
The results of the paper  \cite{Balitsky:2013npa}  have been important to us not only for the appropriate definition of analytic continuation of local operators \eqref{twist2op} to the complex spins \(j\) but also to normalize the three-point function and extract the corresponding structure constants.

A generic configuration of such  3 light-ray operators
in the coordinate space leads to a very complicated, albeit predictable from the group theory,  tensor structure of the 3-point correlator: even for an integer spin \(j\) there will be a few tensor structures, each with its own structure constant.  If we analytically continue to the complex spins it is not even clear how to handle all these structures. That why we have chosen a particular spacial configuration where all three operators are placed to the same line in the transverse two-dimensional subspace.  In this case, after integration,  only one tensor structure is left. Moreover,  it admits a reasonable analytic continuation to the complex spins.\footnote{Note however  that after a conformal transformation the three points in the transverse space take arbitrary positions.
}

To fix the structure constant in front of this remaining tensor structure in the BFKL limit we perform an independent calculation of the 3-point function using the BFKL and BK evolution technique.  We observe the same tensor structure as predicted from the group theoretical arguments \cite{Costa:2011mg}  with the explicit structure constant as a function of BFKL parameters \(\frac{g^2}{\omega_j}\,,\,\,\, j=1,2,3\).

\section{Light-ray  operators and their conformal correlators   }

In this section, we will repeat the definition of twist-2 light-ray operators of N=4 SYM     which have been
proven very useful for the calculation of two-point correlators in \cite{Balitsky:2013npa}. First, we introduce the bi-local
fields with the adjoint Wilson lines  connecting the two local fields, and second, we integrate the positions of these fields along the light-ray with
a  weight which is a power of the distance between them. The short distance
expansion  of this object generates the local   twist-2 operators and their  descendants of higher conformal spin.
In that sense,  a light-ray operator is a generating function of local  twist-2 operators,  but their advantage is a possibility of a natural analytic continuation  w.r.t. conformal spin.

The twist-2 supermultiplet of  local operators was explicitly constructed in \cite{Belitsky:2003sh}. For example, the component with the lowest anomalous dimension at zero order in the SYM coupling \(g_{{}_{YM}}\)  reads as follows (\(j\) is even):
\begin{equation}
\mathcal{S}_{\rm loc}^{j}(x)=6\mathcal{O}_{gg}^{j}(x)+\frac{j-1}{2}
\mathcal{O}_{qq}^{j}(x)+\frac{j(j-1)}{4}\mathcal{O}_{ss}^{j}(x),\label{LocOper}
\end{equation}
where
\begin{eqnarray}
\mathcal{O}_{gg}^{j}(x)&=& \sigma_j\tr \mathcal{G}^{\frac{5}{2}}_{j-2,x_1,x_2} F^{\ \ \mu}_{+\bot}(x_1)g^{\bot}_{\mu\nu}F^{\ \ \nu}_{+\bot}(x_2)|_{x=x_1=x_2},\\
\mathcal{O}_{qq}^{j}(x)&=& \sigma_j \tr \mathcal{G}^{\frac{3}{2}}_{j-1,x_1,x_2} \bar{\lambda}_{\dot{\alpha}A}\sigma^{+\dot{\alpha}\beta}(x_1)\lambda^A_\beta(x_2)|_{x=x_1=x_2},\\
\mathcal{O}_{ss}^{j}(x)&=& \sigma_j \tr \mathcal{G}^{\frac{1}{2}}_{j,x_1,x_2} \bar{\phi}_{AB}(x_1)\phi^{AB}(x_2)|_{x=x_1=x_2}.
\end{eqnarray}
We   introduced here  the  differential operator \(\mathcal{G}^{\alpha}_{n,x_1,x_2}=i^n( \nabla_{x_2} +\nabla_{x_1})^n C_n^{\alpha}(\frac{\nabla_{x_2}-\nabla_{x_1}}{\nabla_{x_2}+\nabla_{x_1}})\), where \(C_n^{\alpha}(x)\) is the Gegenbauer polynomial of order \(n\) with index \(\alpha\) and \(\nabla_x\) are covariant derivatives in
the light-like direction \(n_+\): \(\nabla_x=n_+^\mu(\partial_\mu-ig_{{}_{YM}}A_\mu)=\partial_+-ig_{{}_{YM}}A_+\). Signature factor \(\sigma_j\) is defined as \(\sigma_j=(1+(-1)^j)/2\)

The generalization of \(\mathcal{S}_{\rm loc}^{j}\) to the case of complex spin \(j\)  constructed in  \cite{Balitsky:2013npa}  has a form of light-ray operator stretched along \(n_+\) direction and realizing the  principal series irreducible representation of \(sl(2|4)\) with conformal spin \(J=\frac{1}{2}+i\nu\) which is related to \(j\) as \(J=j+1\). Superprimary light-ray twist-2 operators with conformal spin \(J\) and with the lowest anomalous dimension reads as follows:
\begin{gather}\label{FullLR}
\breve{S}^{j+1}(x_{1\bot},x_{3\bot})=-\frac{j(j-1)}{2}\breve{S}_{sc}^{j+1}(x_{1\bot},x_{3\bot})
+i\frac{j-1}{2} \breve{S}_{f}^{j+1}(x_{1\bot},x_{3\bot})+\breve{S}_{gl}^{j+1}(x_{1\bot},x_{3\bot}),
\end{gather}
with   gluon, fermion and scalar terms defined as
\begin{gather}
\breve{S}_{sc}^{j+1}(x_{1\bot})=\int\limits_{-\infty}^{\infty} d x_{1-} \int\limits_{x_{1-}}^{\infty}dx_{2-} (x_{2-}-x_{1-})^{-j-1}\ \tr\bar{\phi}_{AB}(x_{1})[x_1,x_2]_{Adj}\phi^{AB}(x_{2}), \label{nonlocSc}\\
\breve{S}_{f}^{j+1}(x_{1\bot})=\int\limits_{-\infty}^{\infty} d x_{1-} \int\limits_{x_{1-}}^{\infty}dx_{2-}  (x_{2-}-x_{1-})^{-j}\ \tr\bar{\lambda}_{\dot{\alpha}A}(x_{1})\sigma^{+\dot{\alpha}\beta}[x_1,x_2]_{Adj}
 \lambda^A_\beta(x_{2}),\label{nonlocF}\\
\breve{S}_{gl}^{j+1}(x_{1\bot})=\int\limits_{-\infty}^{\infty} d x_{1-} \int\limits_{x_{1-}}^{\infty}dx_{2-} (x_{2-}-x_{1-})^{-j+1}\ \tr F^{\ \ \mu}_{+\bot}(x_{1})g^{\bot}_{\mu\nu}[x_1,x_2]_{Adj}F^{\ \ \nu}_{+\bot}(x_{2}).\label{nonlocGluon}
\end{gather}
where we defined the Wilson line operator in adjoint representation along the straight
line  stretched between two points:
\begin{equation}
[x_1,x_2]_{Adj}=\text{Pexp}\left[ig\int\limits_{0}^{1}d\a A^{-}\left(x+\a(y-x)\right)\right]_{Adj}
\end{equation}
where \(x=(x_{-},0,x_{\bot}),\ \ y=(y_{-},0,y_{\bot})\).

As they are  written, the formulas   \eqref{nonlocSc}-\eqref{nonlocGluon}  are well defined for complex \(j\). For positive even integer \(j\)'s, the
second integral should be understood as a contour integral  along a contour
going along the
real axis  from \(+\infty\), then  surrounding the pole at \(x_{2-}=x_{1-}\)  and going back to \(+\infty.\)   In  the case of  even integer Lorentz spin \(j\) these formulas can be rewritten as single integrals of local operator \(\mathcal{O}^{j}(x)\) with dimension \(\Delta(j)\) along their light ray direction \(n_+\). The way to do it is to do the  OPE of bi-local operator  under the trace in the integrand around the point  \(x_{1-}\), i.e.,      w.r.t the distances   \((x_{2-}-x_{1-})\)   until the overall    dependence on the distance  gives
the first order pole \((x_{2-}-x_{1-})^{-1}\). The integral in \(x_{2-}\) can then be computed as the pole residue so only the second integral is left. The light-ray operator reads
\begin{gather}
\breve{\mathcal{S}}^{j+1}(x_{\bot})|_{j\in \mathbb{N}}=\frac{\pi^{\frac{3}{2}}2^{1-2j}}{\Gamma(j+\frac{1}{2})\sin \pi j \cos\frac{\pi j}{2}}\int\limits_{-\infty}^\infty dx_- \mathcal{O}_{\rm loc}^{j}(x_-)
\end{gather}

In the paper  \cite{Balitsky:2013npa} we calculated  the correlator of two light-ray operators separated by the transverse distance \(|x_\bot-y_\bot|\) and stretched along \(n_+\) and \(n_-\) vectors normalized according  to
 \(\langle n_+n_-\rangle\)=1. On the other hand, according to
the above argument,  it reduces just  to the double integral of  two-point correlator of local operators w.r.t.  light-ray directions \(x_\pm\).
Using the  conformal form of correlator of our local operators we obtain:
\begin{gather}\label{2pcorrform}
<\breve{\mathcal{S}}^{j+1}(x_{\bot}) \breve{\mathcal{S}}^{j'+1}(y_{\bot})>
=\frac{\delta(j-j') b_{j}}{(|x_\bot-y_\bot|^2)^{\Delta(j)-1}}
\end{gather} where \(b_{j}\)  is a normalization constant.

In this paper we compute  the  correlator of three light-ray operators restricting ourselves to a particular simple kinematics:  one light-ray operator is stretched along \(n_+\) light-ray direction and two other -- along  \(n_-\).
The conformal form of this correlator  can be established by integrating the correlator of 3 local operators along these light-rays and then making analytical continuation to comlex spins. The tensor structures of such local correlators are known from general group-theoretical considerations  \cite{Costa:2011mg} up to a few  structure constants depending  on the coupling and symmetry charges.
Remarkably, if  the coordinates of all 3 light-rays in the transverse space are restricted to the same line all these structures  collapse into a single one \footnote{A similar phenomenon but in the different kinematics was observed in \cite{Kazakov:2012ar}.}, with a single overall structure constant which we are going to compute.

However, the configuration with    two collinear light-ray operators is singular, so we first consider three different polarizations \(n_1,\ n_2,\ n_3\)  and then take the limit \(n_2\rightarrow n_3\). The result of integration along light-rays is quite simple and contains only one unknown overall constant (see appendix \ref{ApConfStrOf3LR} for details)
\begin{gather}
\langle\breve{\mathcal{S}}^{j_1+1}(x_{\bot})\breve{\mathcal{S}}^{j_2+1}(y_{\bot})\breve{\mathcal{S}}^{j_3+1}(z_{\bot})\rangle=\notag\\
=C_{\{n_i\}}(\{\Delta_i\},\{j_i\})\frac{<n_1n_2>^{[j]_{1,2;3}}<n_1n_3>^{[j]_{1,3;2}}<n_2n_3>^{[j]_{2,3;1}}}
{(|x-y|_{\bot}^2)^{[\Delta]_{1,2;3}}(|x-z|_{\bot}^2)^{[\Delta]_{1,3;2}}(|y-z|_{\bot}^2)^{[\Delta]_{2,3;1}}}
\end{gather}
where we  used a short-hand notation \([a]_{i,j;k}\equiv \frac{1}{2}(a_i+a_j-a_k-1)\) and \(\{a_i\}\equiv\{a_1,a_2,a_3\}\).
In what follows we assume  the existence of a good analytic continuation for \(C_{\{n_i\}}(\{\Delta(j_i)\},\{j_i\})\) to non-integer \(\{j_i\}\)'s. We take the limit \(n_1=n_+,\ n_2=n_-,\ n_3\rightarrow n_2\) with the normalization \(\langle n_+ n_-\rangle=1\). In the BFKL regime \(j_i=1+\omega_i\rightarrow 1\) we obtain:
\begin{gather}
\langle\breve{\mathcal{S}}^{2+\omega_1}(x_{\bot})\breve{\mathcal{S}}^{2+\omega_2}(y_{\bot})\breve{\mathcal{S}}^{2+\omega_3}(z_{\bot})\rangle=\notag\\
=\lim\limits_{n_3\rightarrow n_2=n_-} \frac{<n_2n_3>^{\frac{\omega_2+\omega_3-\omega_1}{2}}}{\omega_2+\omega_3-\omega_1}\frac{C_{+--}(\{\Delta_i\},\{1+\omega_i\})}
{|x-y|_{\bot}^2)^{[\Delta]_{1,2;3}}(|x-z|_{\bot}^2)^{[\Delta]_{1,3;2}}(|y-z|_{\bot}^2)^{[\Delta]_{2,3;1}}}\label{3pFromLocals}
\end{gather}
where \(\Delta_i=\Delta(1+\omega_i,g^2) \)  is given by BFKL spectrum (see below). Notice that the transverse coordinate dependence  is similar to that of the correlator of  three scalar operators.
We  explicitly pulled out   the denominator \(\frac{1}{\omega_2+\omega_3-\omega_1}\) because it will emerge in our forthcoming calculation using the BK evolution. Notice that the numerator in the second line in the limit \(\langle n_2 n_3\rangle\to 0\) goes either to zero, if \(\omega_2+\omega_3-\omega_1>0\), or to infinity, if \(\omega_2+\omega_3-\omega_1<0\).     We interpret  \(\lim\limits_{<n_2n_3>\rightarrow 0}\frac{<n_2n_3>^{\frac{\omega_2+\omega_3-\omega_1}{2}}}{\omega_2+\omega_3-\omega_1}\) as a "half-delta"-function denoted by \(\delta_{>}(\omega_2+\omega_3-\omega_1)\)
which means of course that in all coefficients in front of this object we
have to put \(\omega_2+\omega_3-\omega_1=0\).\footnote{This delta-function is  reflecting the boost invariance, see text before formula (\ref{IntegralOverY})} In addition, we keep all  \(\omega_i\) positive through the paper.
The same object \(\delta_{>}(\omega_2+\omega_3-\omega_1)\) will emerge in our explicit BFKL computations.

Finally, the  structure constant will be  normalized using the corresponding 2-point correlators:
\begin{gather}\label{normC}
C_{\omega_1,\omega_2,\omega_3}=\frac{C_{+--}
(\{\Delta_i\},\{1+\omega_i\})}{\sqrt{b_{1+\omega_1}b_{1+\omega_2}b_{1+\omega_3}}}
\end{gather}   In what follows, we will compute it  using the BFKL and BK evolution.

\section{Decomposition over dipoles and BK evolution}
The light-ray operators defined by formulas  \eqref{nonlocSc}-\eqref{nonlocGluon}  are not well defined due to the singularity occurring when the positions of two fields  on the light-ray coincide.  In calculation  the two-point correlator  \cite{Balitsky:2013npa} we  used  a point splitting regularization in orthogonal direction. We replace
 light-rays by infinitely narrow Wilson frames with inserted fields in the corners (see fig.\ref{fig:2p}). To regularize the gluon light-ray operator which dominates in the BFKL limit we also proposed a simpler regularization, replacing the direct insertions of \(F_{\mu\nu}\) fields by derivatives w.r.t. the corner coordinates of a pure Wilson frame.  For the  sake of simplicity, we carry out our calculation for the pure Wilson frames, related to our operators with zero \(R\)-charge in the following way:
\begin{gather}
\partial_{x_{1\bot }}\cdot\partial_{x_{3\bot}}
 \int \int\frac{dx_{1-}dx_{3-}}{(x_{3-}-x_{1-})^{2+\omega}}
\,\,[x_1,x_3]_{\Box_\pm}
\underset{x_{13\bot}\rightarrow 0,\,\omega\rightarrow 0}{\rightarrow}
 |x_{13\bot}|^{\gamma_\omega}c(g_{YM}^2,N_c,\omega)\mathcal{S}_\pm^{2+\omega}(x_{1\bot}).
 \label{LightRayFromRamka}
\end{gather}    We  introduced the notation \([x_1,x_3]_{\Box_{\pm}}\) for  rectangular Wilson contour stretched along \(n_\pm\) light-cone directions with coordinates \(x_1,x_3\) of two diagonally opposite corners shown in  Fig.\ref{fig:2p}.

The two definitions should be the same modulo local renormalization of such a light-ray operator, so when calculating the structure constants we should use the two-point functions of light-ray operators with the same definitions for the overall normalization. The   coefficient \(c(g_{YM}^2,N_c,\omega)\) (denoted below as \(c(\omega))\)  depends on the local regularization procedure and at weak coupling it behaves as \(c(\omega)\approx\frac{2g^2_{YM}}{\omega}\), but its explicit form is irrelevant for us because we are going to calculate the normalized structure constant where it cancels.

In general, there are a few types of leading twist-2 operators which appear in the decomposition of left hand side of (\ref{LightRayFromRamka})  but in the BFKL limit a single one with the smallest anomalous dimension survives. In addition, as was argued in ~\cite{Balitsky:2013npa},  in the \(\omega\to 0 \) limit only the term  built out of gauge fields alone  contributes, due to the factors \(j-1\) in the other two terms in \eqref{FullLR}.

Following the OPE method \cite{Balitsky:1995ub},       the pure Wilson frame
(e.g. stretched along \(n_+\)) can be replaced by regularized color dipoles. Before do it let us remind the logic of this approach within the scattering theory and then relate it to our calculation. The high-energy behavior of the amplitudes can be studied in the framework of the rapidity evolution of Wilson-line operators forming color dipoles.
The main idea is the factorization in rapidity: we separate a typical functional integral describing scattering of two particles into (i) the integral over the
gluon fields with rapidity close to the rapidity of the "probe" $Y_A$ , (ii) the integral over the gluons with rapidity close to the rapidity
of the target \(Y_B\), and (iii) the integral over the intermediate region of rapidities \(Y_A>Y>Y_B\)
  (as in Fig. \ref{fig:2p}). The result of the first integration is a certain coefficient function (impact factor) times color dipole (ordered in the
 direction of the probe velocity) with rapidities up to $Y_A$. Similarly, the result of the second integration is again the impact factor
 times the color dipole ordered in the direction of targetХ velocity with rapidities greater than $Y_B$. The result of the last integration is the correlation function
 of two dipoles which can be calculated using the evolution equation for color dipoles,  known in the leading and next-to-leading order \cite{Balitsky:1995ub}.

To factorize in rapidity,  it is convenient to use the background field formalism: we integrate over gluons with $Y>Y_A$ and leave the gluons with $Y<Y_A$ as a background field, to
be integrated over later.  Since the rapidities of  background
gluons are very different from the rapidities of gluons in our Feynman diagrams, the background field is seen by the probe in the form of a shock wave (pancake) due to the Lorentz contraction.
To derive the expression of a quark or gluon propagator in this shock-wave background we represent the propagator as a path integral over various trajectories,
each of them weighed with the gauge factor Pexp$(ig\int\! dx_\mu A^\mu)$ ordered along the propagation path. Now, since the shock wave is very thin, quarks or gluons emitted by the probe do not
have time to deviate in transverse direction so their trajectory inside the shock wave can be approximated by a segment of the straight line. Moreover, since there is no external field
outside the shock wave, the integral over the segment of straight line can be formally extended to \(\pm\infty\) limits yielding the Wilson-line
gauge factor
\begin{gather}
U^{\sigma_+}_{x_\bot}=P\exp[i g_{{}_{YM}}
\int\limits_{-\infty}^\infty dx_+A^{\sigma_+}_-(x)],\\
A_{\mu}^{\sigma_+}(x)=\int d^4k \theta(\sigma_+-|k_+|)e^{ikx}A_\mu(k).
\label{fund line factor}\end{gather}
where we have used the gauge field with a cutoff  \(\s=e^Y\) w.r.t. the longitudinal momenta \(k_+\)

In our case, the Wilson frames play the role of the probe and the target, respectively. The gluons emitted and absorbed within each frame contribute to their   "impact factors". The correlator factorizes into these three impact factors and the BK evolution  of colour dipoles  appearing in the OPE of the frames.      The BFKL evolution corresponds to the evolution of first dipole with rapidity \(Y_1\)  to some intermediate value \(Y_0\). Then nonlinear term in BK evolution produces two dipoles which should be paired with two dipoles corresponding to frames oriented in \(n_-\) direction. This contraction of two dipoles can be made by making use of BFKL evolution as we did in \cite{Balitsky:2013npa}.

Impact factor for pure Wilson frames is trivial due to the fact that one can just replace frames by dipoles with propriety cutoff:
\begin{gather}
[x_1,x_3]_{\Box_+} \rightarrow N(1-\mathbf{U}_+^\sigma(x_{1\bot},x_{3\bot}))
\end{gather}
where
\begin{gather}
\mathbf{U}_+^\sigma(x_{1\bot},x_{3\bot})=
1-\frac{1}{N}\text{tr}\left(U^{\sigma_+}_{x_{1\bot}}(U^{\sigma_{+}}_{x_{3\bot}})^\dagger\right)
\end{gather}
and \(\sigma_+ \) is a rapidity  cutoff (see   \cite{Balitsky:1997mk,Kovchegov:1999ua,Kovchegov:1999yj}
for explanations on the role of cutoff).

Now we can move on to the main goal of our present paper -- the computation
of  correlator of three light-ray operators  defined from the Wilson frames
by \eqref{LightRayFromRamka}:
\begin{gather}
\langle \mathcal{S}_-^{2+\omega_1}(x_{1\bot},x_{3\bot})\mathcal{S}_+^{2+\omega_2}(y_{1\bot},y_{3\bot})\mathcal{S}_+^{2+\omega_3}(z_{1\bot},z_{3\bot})\rangle=\notag\\
={\cal -D}_{\bot}\int\limits_{-\infty}^{\infty}dx_{1-} \int\limits_{x_{1-}}^{\infty}dx_{3-}x_{31-}^{-2-\omega_1} \int\limits_{-\infty}^{\infty}dy_{1+} \int\limits_{y_{1+}}^{\infty}dy_{3+}y_{31+}^{-2-\omega_2}
 \int\limits_{-\infty}^{\infty}dz_{1+} \int\limits_{z_{1+}}^{\infty}dz_{3+}z_{31+}^{-2-\omega_3}\times \notag \\
\times\langle \mathbf{U}^{\sigma_{1-}}(x_{1\bot},x_{3\bot})\mathbf{V}^{\sigma_{2+}}(y_{1\bot},y_{3\bot})\mathbf{W}^{\sigma_{3+}}(z_{1\bot},z_{3\bot}) \rangle,\label{3pStart}
\end{gather}
where, according to \eqref{LightRayFromRamka},  \({\cal D}_{\bot}=\frac{N^3(\partial_{x_{1\bot}}\cdot\partial_{x_{3\bot}})(\partial_{y_{1\bot}}
\cdot\partial_{y_{3\bot}})(
\partial_{z_{1\bot}}\cdot\partial_{z_{3\bot}})}{c(\omega_1)c(\omega_2)c(\omega_3)}\).

In our kinematics, the color dipole \(\mathbf{U}\) is stretched along \(n_+\)
direction and   two dipoles \(\mathbf{V}\) and \(\mathbf{W}\) have zero \(n_+\) projection. In the LO  BFKL approximation the last two form a "pancake" field configuration in the reference frame related to \(\mathbf{U}\). This means that the rapidity of \(\mathbf{U}\) serves as the upper limit (cutoff) for  integrations w.r.t. rapidities of     \(\mathbf{V}\) and \(\mathbf{W}\)  in our logarithmic approximation. Now we use the BK evolution equation
  \cite{Balitsky:1997mk,Kovchegov:1999ua,Kovchegov:1999yj} to calculate the
 quantum average in (\ref{3pStart}). It gives the evolution of the  dipole \(\mathbf{U}^{Y_1}\) with respect to rapidity \(Y_1=e^{\sigma_1}\), namely
\begin{gather}
\sigma \frac{d}{d\sigma}\mathbf{U}^\sigma(z_1,z_2)=\mathcal{K}_{{\rm BK}}\ast \mathbf{U}^\sigma(z_1,z_2), \label{BK}
\end{gather}
whereis a non-linear integral operator  \(\mathcal{K}_{\rm BK}\)  has the following form in LO  approximation:
\begin{gather}
\mathcal{K}_{_{\rm LO\, BK}}\ast\mathbf{U}(z_1,z_2)=\frac{2g^2}{\pi} \int d^2z_3\frac{z_{12}^2}{z_{13}^2z_{23}^2}\left[\mathbf{U}(z_1,z_3)+\right.\notag\\
\left. +\mathbf{U}(z_3,z_2)-\mathbf{U}(z_1,z_2)-\mathbf{U}(z_1,z_3)\mathbf{U}(z_3,z_2)\right].
\end{gather}
 The result of the evolution of color dipole \(U\) from \(Y_1$ to \(\max (Y_2,Y_3)\) (suppose it is \(Y_2\)) can be formally written as
\footnote
{In principle, one can attach the color dipole in (+) direction to
two color dipoles in (-) direction by exchange of two pomerons.
It can be demonstrated, however, that such contribution behaves like \(z_{12}^4\) and corresponds to the LR operators
of higher twists in comparison to  twist-two terms \(\sim z_{12}^2\) arising as a result of the evolution (\ref{3.9}).
}
\begin{equation}
e^{(Y_1-Y_2)\mathcal{K}_{\rm BK}}\ast \mathbf{U}(z_1,z_2)~=~e^{(Y_1-Y_2)(\mathcal{K}_{\rm BFKL}+\mathcal{K}_{\rm NLBK})}\ast \mathbf{U}(z_1,z_2)
\label{3.9}
\end{equation}
where the \(\mathcal{K}_{\rm BFKL}\) is the linear BFKL part of the kernel (3.8) and \(\mathcal{K}_{\rm NLBK}\) is the non-linear term.
In general, the result of the evolution of a color dipole at rapidity \(Y_1\) is a multitude of color dipoles (and other color
structures like quadrupoles) at rapidity \(Y_2\). However, each non-linear term in the r.h.s. of Eq. (3.8) brings an extra \(g^2\) in comparison to linear ones,
so we can restrict ourselves to minimal number of non-linear terms in the evolution operator
\(e^{(Y_1-Y_2)\mathcal{K}_{\rm BK}}\). So we can expand the r.h.s. w.r.t. the non-linear term up to the first order. This gives
\begin{equation}
e^{(Y_1-Y_2)\mathcal{K}_{\rm BK}}\ast \mathbf{U}(z_1,z_2)\,=\int_{Y_2}^{Y_1} dY_0\,\,
e^{(Y_1-Y_0)\mathcal{K}_{\rm BFKL}}\,\mathcal{K}_{\rm NLBK}\,\,
e^{(Y_0-Y_2)(\mathcal{K}^{(1)}_{\rm BFKL}+\mathcal{K}^{(2)}_{\rm BFKL})}\ast \mathbf{U}(z_1,z_2)
\label{3.10}
\end{equation}

\begin{figure}
  \centering
  \includegraphics[scale=0.8]{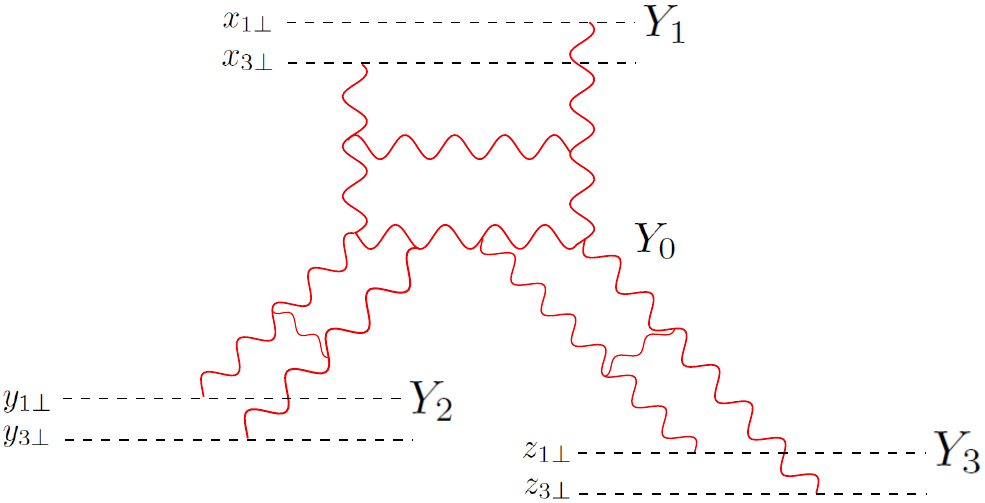}
  \caption{3 pomeron contribution corresponding to the calculation of 3-point correlator}
  \label{fig:3pomeron}
\end{figure}

Evolution of \(\mathbf{U}^{Y_1}\) goes from \(Y_1\) to an intermediate \(Y_0\) w.r.t. the linear part of (\ref{BK}), and then the BK vertex   acts at  \(Y_0\) and generates two dipoles evolving with kernels \(\mathcal{K}^{(1)}_{\rm BFKL}\) and
\(\mathcal{K}^{(2)}_{\rm BFKL}\). Finally, these two  dipoles  \(\mathbf{U}^{Y_{2}}\) can be contracted with \(\mathbf{V}^{Y_{2}}\) and \(\mathbf{W}^{Y_{3}}\).
Schematically, it can be written as:
\begin{gather*}
\int dY_0(\mathbf{U}^{Y_{1}}\rightarrow \mathbf{U}^{Y_0})\otimes(\text{BK\ vertex at}\,\,Y_0 )\otimes\begin{pmatrix}\langle \mathbf{U}^{Y_{0}} \mathbf{V}^{Y_{2}}\rangle \\
\langle \mathbf{U}^{Y_{0}} \mathbf{W}^{Y_{3}}\rangle \\
\end{pmatrix}.
\end{gather*}
and corresponding diagram is presented in  Fig.\ref{fig:3pomeron}

The linear BFKL evolution of \(\mathbf{U}^{Y_1}\) from \(Y_1\) to \(Y_0\) gives:
\begin{gather}
\mathbf{U}^{Y_1}(x_1,x_3)=\int d\nu\int d^2x_0 \frac{\nu_1^2}{\pi^2}E_{\nu_1}(x_{10},x_{30})e^{\aleph(\nu_1)Y_{10}}\cdot\notag\\
\cdot\frac{1}{\pi^2}\int
\frac{d^2\gamma d^2\beta}{|\gamma-\beta|^4}E^{*}_{\nu_1}(\gamma-x_0,\beta-x_0)\mathbf{U}^{Y_0}(\gamma,\beta),
\end{gather}
where we denoted \(Y_{ij}\equiv Y_i-Y_j\) and we  introduced the function
\begin{gather}
E_\nu(z_{10},z_{20})=\left(\frac{|z_{12}|^2}{|z_{10}|^2|z_{20}|^2}\right)^{1/2+i\nu},
\end{gather}
which projects dipoles on the eigenstates of BFKL operator with the eigenvalues
\begin{gather}
\aleph(\nu)=4g^2(2\psi(1)-\psi(1/2+i\nu)-\psi(1/2-i\nu)).
\end{gather}
We take here only the sector  \(n=0\), where \(n\) is the discrete quantum number of \(SL(2,C)\) because it gives the leading contribution.

The non-linear part of BK evolution   (\ref{BK}) is described by the following equation:
\begin{gather}
\left.\frac{\partial}{\partial Y}\mathbf{U}^{Y}(\gamma,\beta)\right|_{Y=Y_0}=-\frac{2g^2}{\pi}
 \int d^2\alpha\frac{|\gamma-\beta|^2}{|\gamma-\alpha|^2|\beta-\alpha|^2}\mathbf{U}^{Y_0}(\gamma,\alpha)\mathbf{U}^{Y_0}(\alpha,\beta)
\end{gather}
\section{Computation of 3-point correlator}
Finally, we  contract the two emerging dipoles \(\mathbf{U}^{Y_0}(\gamma,\alpha)\) and \(\mathbf{U}^{Y_0}(\alpha,\beta)\) with \(\mathbf{V}^{\sigma_{2+}}(y_{1\bot},y_{3\bot})\) and \(\mathbf{W}^{\sigma_{3+}}(z_{1\bot},z_{3\bot})\). Thus for the planar contribution we get:
\begin{gather}
\langle \mathbf{U}^{Y_{1}}(x_{1\bot},x_{3\bot})\mathbf{V}^{Y_{2}}(y_{1\bot},y_{3\bot})\mathbf{W}^{Y_{3}}(z_{1\bot},z_{3\bot}) \rangle_{pl}=\label{evolutionUpDown}\\
=-\frac{2g^2}{\pi}\int d Y_0 \int d\nu_1\int d^2x_0 \frac{\nu_1^2}{\pi^2}E_{\nu_1}(x_{10},x_{30})e^{\aleph(\nu_1)Y_{10}}\times\notag\\
\times\frac{1}{\pi^2}\int
\frac{d^2\alpha d^2\beta d^2\gamma}{|\gamma-\beta|^2 |\gamma-\alpha|^2|\beta-\alpha|^2}E^{*}_{\nu_1}(\gamma-x_0,\beta-x_0)\cdot \notag\\
\cdot  (\langle \mathbf{U}^{Y_0}(\gamma,\alpha)\mathbf{V}^{Y_{2}}(y_{1\bot},y_{3\bot})\rangle \langle\mathbf{U}^{Y_0}(\alpha,\beta) \mathbf{W}^{Y_{3}}(z_{1\bot},z_{3\bot}) \rangle +\notag\\
+\langle \mathbf{U}^{Y_0}(\gamma,\alpha)\mathbf{W}^{Y_{3}}(z_{1\bot},z_{3\bot})\rangle \langle\mathbf{U}^{Y_0}(\alpha,\beta) \mathbf{V}^{Y_{2}}(y_{1\bot},y_{3\bot}) \rangle)\notag
\end{gather}
The last two terms in (\ref{evolutionUpDown}) give the same contribution so it is enough  to know the correlators of two dipoles \cite{Balitsky:2013npa}:
\begin{gather}
\langle \mathbf{U}^{Y_0}(\gamma,\alpha)\mathbf{V}^{Y_{2}}(y_{1\bot},y_{3\bot})\rangle=\notag\\
=\frac{8g^4(1-N_c^2)}{N_c^4}\int d^2y_0 \int\frac{d\nu_2 \nu_2^2 e^{Y_{02}\aleph(\nu_2)}}{(\frac{1}{4}+\nu_2^2)^2} E_{\nu_2}(\gamma-y_0,\alpha-y_0) E^*_{\nu_2}(y_{10},y_{30}) \\
 \langle\mathbf{U}^{Y_0}(\alpha,\beta) \mathbf{W}^{Y_{3}}(z_{1\bot},z_{3\bot}) =\notag\\
=\frac{8g^4(1-N_c^2)}{N_c^4}\int d^2z_0 \int\frac{d\nu_3 \nu_3^2 e^{Y_{03}\aleph(\nu_3)}}{(\frac{1}{4}+\nu_3^2)^2} E_{\nu_3}(\gamma-z_0,\alpha-z_0) E^*_{\nu_3}(z_{10},z_{30})  \label{NizhnyProp2}
\end{gather}
It was argued in  \cite{Balitsky:2013npa} that in case of two-point correlator we can choose the cutoff using anharmonic ratios \cite{Cornalba:2007fs}:
\begin{gather}
e^{Y_{12}\aleph(\nu)} \rightarrow \notag\\
\rightarrow\frac{-i}{\sin \pi \aleph(\nu)}
\left(\frac{((x_1-y_3)^2)^{\frac{\aleph(\nu)}{2}}((x_{3}-y_1)^2)^{\frac{\aleph(\nu)}{2}}}{(x_{13}^2)^{\frac{\aleph(\nu)}{2}}(y_{13}^2)^{\frac{\aleph(\nu)}{2}}}-
\frac{((x_1-y_1)^2)^{\frac{\aleph(\nu)}{2}}
((x_{3}-y_3)^2)^{\frac{\aleph(\nu)}{2}}}{(x_{13}^2)^{\frac{\aleph(\nu)}{2}}(y_{13}^2)^{\frac{\aleph(\nu)}{2}}}\right).
\label{cutoff}
\end{gather}
In the LO approximation we can take just an asymptotic:
\begin{gather}
e^{Y_{12}\aleph(\nu)} \rightarrow \left(\frac{x_{31-}y_{31+}}{\Lambda^2}\right)^{\aleph(\nu)},
\end{gather}
where \(\Lambda\) is a cutoff whose precise value is irrelevant for us. Using this identification and introducing \(L_0\) for the intermediate rapidity \(Y_0=\log \frac{L_0}{\Lambda}\) we can identify all rapidities in the following way:
\begin{gather}
Y_{10}=\log \frac{x_{31-}}{L_0}  , \ \ \ Y_{02}=\log\frac{L_0 y_{31+}}{\Lambda^2}, \ \ Y_{03}=\log\frac{L_0 z_{31+}}{\Lambda^2}
\end{gather}
Integral over rapidities reads as:
\begin{gather}
\int L_1^{-1-\omega_1} \int L_2^{-1-\omega_2} \int L_3^{-1-\omega_3} \int d Y_0 e^{Y_{10}\aleph_1+Y_{02}\aleph_2+Y_{03} \aleph_3} \theta (Y_{10}) \theta(Y_0-max(Y_2,Y_3))
\end{gather}

This integral diverges and we should introduce a proper regularization. In order to compare our BFKL calculation with (\ref{3pFromLocals}) let's slightly change the third polarization \(n_3\) from \(n_-\) and then take a limit \(n_3\rightarrow n_-\). We can use our  formulas for $n_2=n_3$ case until when
$y_{31+}z_{31+}(n_2\cdot n_3)\leq  \Delta_\perp^2$ where \(\Delta_\perp\) is a typical scale in orthogonal direction. Again its precise value doesn't matter in LO BFKL. In terms of rapidities this restriction means
$Y_2+Y_3 \leq \log \frac{1}{n_2\cdot n_3}$ and leads to one extra \(\theta\)-function:
\begin{gather}
\int\int\int dY_1dY_2dY_3 dY_0~\theta(Y_1-Y_0)\theta(Y_0+Y_2)\theta(Y_0+Y_3)\theta\big(\log \frac{1}{n_2\cdot n_3}-Y_2-Y_3)\cdot\notag\\
\cdot e^{-\omega_1Y_1-\omega_2Y_2-\omega_3Y_3+\aleph_1(Y_1-Y_0)+\aleph_2(Y_0+Y_2)+\aleph_3(Y_0+Y_3)}=
\notag\\
=\frac{(n_2 \cdot n_3)^{\frac{\omega_2+\omega_3-\omega_1}{2}}}{\omega_1-\omega_2-\omega_3} \frac{1}{(\omega_1-\aleph_1)(\omega_2-\aleph_2)
(\omega_3-\aleph_3)} \label{IntegralOverY}
\end{gather}

As in (\ref{3pFromLocals}) the first factor can be interpreted in the limit \(n_3\rightarrow n_2=n_-\) as a half-delta-function and we can establish relation between our BFKL calculation and analytical continuation of structure constant:

After this integration over configurations of Wilson frames only the coordinate integrals over the two-dimensional orthogonal space are left:
\begin{gather}
\langle S^{2+\omega_1}(x_{1\bot},x_{3\bot})S^{2+\omega_2}(y_{1\bot},y_{3\bot})S^{2+\omega_3}(z_{1\bot},z_{3\bot})\rangle_{pl}=\notag\\
=\frac{2^8 g^{10}(N_c^2-1)^2}{\pi^3 N_c^8}\delta_>(\omega_1-\omega_2-\omega_3)\cdot\notag\\
\cdot D_{\bot}\int d\nu_1\frac{\nu_1^2}{\pi^2}\frac{1}{\omega_2+\omega_3-\aleph(\nu_1)} \int \frac{d\nu_2 \nu_2^2}{(\frac{1}{4}+\nu_2^2)^2} \frac{1}{\omega_2-\aleph(\nu_2)} \cdot\notag \\ \cdot \int  \frac{d\nu_3 \nu_3^2}{(\frac{1}{4}+\nu_3^2)^2}\frac{1}{\omega_3-\aleph(\nu_3)} \int d^2x_0 d^2y_0 d^2z_0  E^*_{\nu_1}(x_{10},x_{30})\cdot\notag\\
\cdot E^*_{\nu_2}(y_{10},y_{30}) E^*_{\nu_3}(z_{10},z_{30})\Upsilon_{pl} (\nu_1,\nu_2,\nu_3;x_0,y_0,z_0) \label{Ramki3pConfRepPlanar}
\end{gather}
where \(\Upsilon_{pl}\) represents the planar contribution of BK vertex:
\begin{gather}
\Upsilon_{pl} (\nu_1,\nu_2,\nu_3;x_0,y_0,z_0)=\notag\\
=\int\frac{d^2\alpha d^2\beta d^2\gamma}{|\gamma-\beta|^2 |\gamma-\alpha|^2|\beta-\alpha|^2}
E_{\nu_1}(\beta-x_0,\gamma-x_0)E_{\nu_2}(\alpha-y_0,\gamma-y_0)E_{\nu_3}(\alpha-z_0,\beta-z_0)=\notag\\
=\frac{\Omega(h_1,h_2,h_3)}
{|x_0-y_0|^{4[h]_{1,2;3}+2}\ |x_0-z_0|^{4[h]_{1,3;2}+2}\ |y_0-z_0|^{4[h]_{2,3;1}+2}}\label{Upsilon_pl}
\end{gather}
Here \(h_1=\frac{1}{2}+i\nu_1,\,h_2=\frac{1}{2}+i\nu_2,\,h_3=\frac{1}{2}+i\nu_3\) and the function \(\Omega(h_1,h_2,h_3)\) was presented in  \cite{Korchemsky:1997fy}.

\begin{figure}
  \centering
  \includegraphics[scale=0.33]{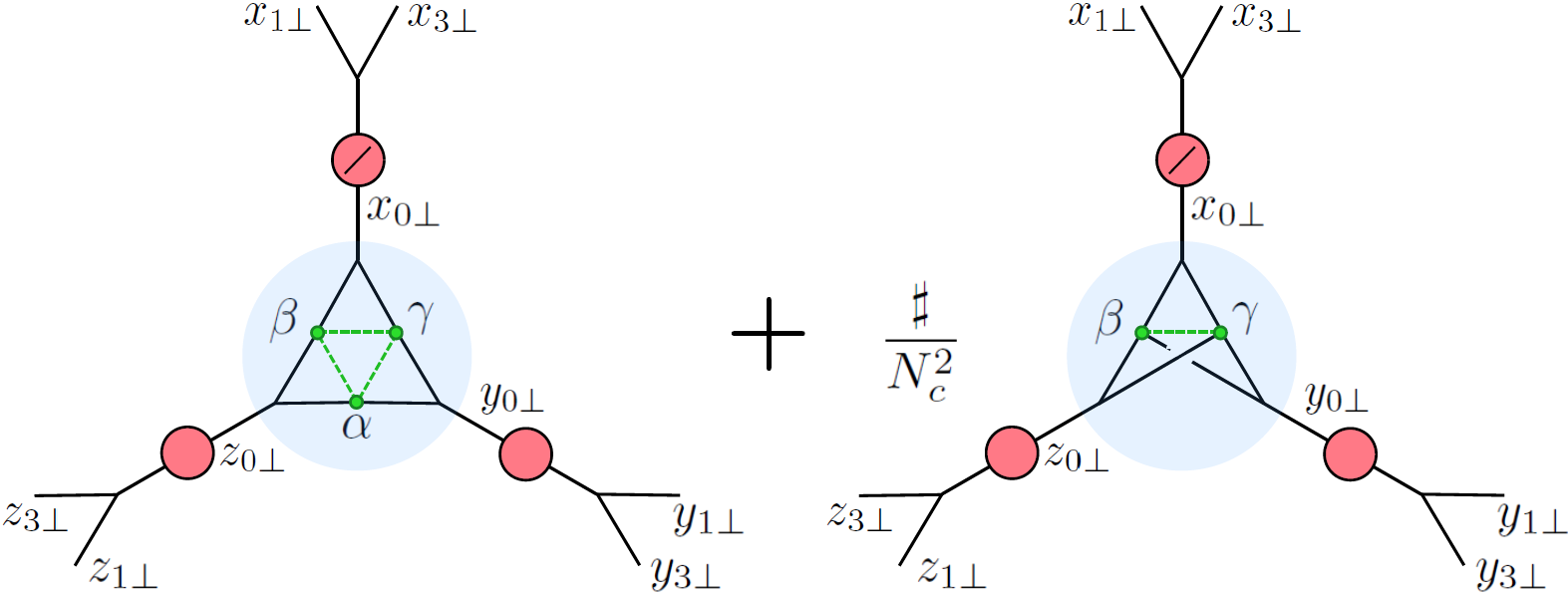}
  \caption{The structure of 3-point correlator. Red circles correspond to BFKL propagators (the crossed one has extra multiplier \((\frac{1}{4}+\nu_1^2)^2\)). The blue blob  corresponds to the  3-point functions of 2-dimensional  BFKL CFT.  The triple veritces correspond to \(E\)-functions. The \(\alpha\beta\gamma\)-triangle in the first, planar, term and \(\beta\gamma\)-link in the second, nonplanar, term correspond to  triple pomeron vertex.}
  \label{fig:3p}
\end{figure}

Remarkably, we can also take into account the non-planar contribution ,   thus providing the finite \(N_c\) answer for the BFKL structure constant! It appears \cite{Korchemsky:1997fy,Chirilli:2010mw} as a single extra term \(\Upsilon_{npl}\):
\begin{gather}
\Upsilon_{npl} (\nu_1,\nu_2,\nu_3;x_0,y_0,z_0)
=\notag\\
=\int\frac{d^2\beta d^2\gamma}{|\gamma-\beta|^4} E_{\nu_1}(\beta-x_0,
\gamma-x_0)E_{\nu_2}(\beta-y_0,\gamma-y_0)E_{\nu_3}(\beta-z_0,\gamma-z_0)=\notag\\
=\frac{\Lambda(h_1,h_2,h_3)}
{|x_0-y_0|^{4[h]_{1,2;3}+2}\ |x_0-z_0|^{4[h]_{1,3;2}+2}\ |y_0-z_0|^{4[h]_{2,3;1}+2}} \label{Upsilon_npl}
\end{gather}
where  \(\Lambda(h_1,h_2,h_3)\) was also presented in  \cite{Korchemsky:1997fy}, and the full answer can be obtained from (\ref{Ramki3pConfRepPlanar}) by replacing \(\Upsilon_{pl}\) with \(\Upsilon\) (see in \ref{fig:3p}):
\begin{gather}
\Upsilon=\Upsilon_{pl}-\frac{2\pi}{N^2}\Upsilon_{npl} Re [ \psi(1)+\psi(\frac{1}{2}+i\nu_1)-\psi(\frac{1}{2}+i\nu_2)-\psi(\frac{1}{2}+i\nu_3)].\label{Upsilon}
\end{gather}
The integrals over \(x_0,y_0,z_0\) are easily computable, e.g.
\begin{gather}
\int d^2x_0 E_{\nu_1}(\beta-x_0,\gamma-x_0)E^*_{\nu_1}(x_{10},x_{30})=\notag\\
=(\tau^2)^{\frac{1}{2}+i\nu_1}
{}_2F_1(\frac{1}{2}+i\nu,\frac{1}{2}+i\nu,1+2i\nu,\tau) {}_2F_1(\frac{1}{2}+i\nu,\frac{1}{2}+i\nu,1+2i\nu,\bar{\tau})\frac{(\frac{1}{4}+\nu^2)^2}{\nu^2}G(\nu)+\notag\\
+(\nu\leftrightarrow -\nu),\label{Lipatov4p}\\
G(\nu)=\frac{\nu^2}{(\frac{1}{4}+\nu^2)^2}\frac{\pi \Gamma^2(\frac{1}{2}+i\nu)\Gamma(-2i\nu)}{\Gamma^2(\frac{1}{2}-i\nu)\Gamma(1+2i\nu)},
\end{gather}
where \(\tau=\frac{|x_1-x_3||\beta-\gamma|}{|x_1-\beta||x_3-\gamma|}\). In the limit \(x_1,x_3\rightarrow x\) we can replace (see appendix \ref{AroundConvPoints} for details):
\begin{gather}
\frac{|x_1-x_3||\beta-\gamma|}{|x_1-\beta||x_3-\gamma|}\rightarrow\frac{|x_1-x_3||\beta-\gamma|}{|x-\beta||x-\gamma|}\label{replacement}
\end{gather}.
For small \(\tau\) we  close the  \(\nu_1\) contour in the lower (upper) half-plane for first(second) term, respectively, both of them giving the same contribution. Integrals over \(\alpha,\beta,\gamma\)  in \eqref{Ramki3pConfRepPlanar} can be  reduced to   \(\Upsilon_{pl}\)   represented in \cite{Korchemsky:1997fy} in terms of   hypergeometric and Meijer G functions, and \(\Upsilon_{npl}\)   in terms of \(\Gamma\)-functions. Integrals over \(\nu_i\) can be done by picking up the BFKL poles \(\omega_i=\aleph(\nu^*_i)\).

Combining (\ref{Ramki3pConfRepPlanar}),(\ref{Upsilon}) and (\ref{Lipatov4p}) we come to the final expression for 3-point correlation function:
\begin{gather}
\langle \mathcal{S}^{2+\omega_1}(x_{1\bot},x_{3\bot})\mathcal{S}^{2+\omega_2}(y_{1\bot},y_{3\bot})\mathcal{S}^{2+\omega_3}(z_{1\bot},z_{3\bot})\rangle=\notag\\
=-i g^{10} \frac{\delta(\omega_1-\omega_2-\omega_3)}{c(\omega_1)c(\omega_2)c(\omega_3)}H\frac{\Psi(\nu_1^*,\nu_2^*,\nu_3^*)|x_{13}|^{\gamma_1}|y_{13}|^{\gamma_2}|z_{13}|^{\gamma_3}}
{|x-y|^{2+\gamma_1+\gamma_2-\gamma_3}|x-z|^{2+\gamma_1+\gamma_3-\gamma_2}|y-z|^{2+\gamma_2+\gamma_3-\gamma_1}}\label{Otvet}
\end{gather}
where
\begin{gather}
\quad H=\frac{2^{10}(N_c^2-1)^2}{\pi^2 N_c^5}\gamma_1^2(2+\gamma_1)^4(2+\gamma_2)^2(2+\gamma_3)^2\frac{G(\nu_1^*)}{\aleph'(\nu_1^*)} \frac{G(\nu_2^*)}{\aleph'(\nu_2^*)} \frac{G(\nu_3^*)}{\aleph'(\nu_3^*)},
\label{3pointfinal}
\end{gather}
\(\gamma_i=\gamma(1+\omega_i)\) - anomalous dimension and the coefficient \(\Psi(\nu_1^*,\nu_2^*,\nu_3^*)\) can be expressed through the  functions \(\Omega(h_1,h_2,h_3)\)  and \(\Lambda(h_1,h_2,h_3)\) defined in (\ref{Upsilon_pl})-(\ref{Upsilon_npl}) and calculated in \cite{Korchemsky:1997fy}:
\begin{gather}
\Psi(\nu_1^*,\nu_2^*,\nu_3^*)=\Omega(h^*_1,h^*_2,h^*_3)-\frac{2\pi}{N_c^2}\Lambda(h^*_1,h^*_2,h^*_3)\text{Re}(\psi(1)-\psi(h^*_1)-\psi(h^*_2)-\psi(h^*_3)),
\end{gather}
where \(h^*_i=\frac{1}{2}+i\nu_i^*=1+\frac{\gamma_i}{2}\).

In order to calculate normalized structure constant we need \(b_{1+\omega}\) coefficient from (\ref{2pcorrform}). It can be calculated as in \cite{Balitsky:2013npa} and result reads as:
\begin{gather}
b_{1+\omega}=-i 2^6 \frac{g^4}{c_\omega^2} (1-\frac{1}{N_c^2})\pi^2 (2+\gamma)^4 \frac{G(\nu^*)}{\aleph'(\nu^*)}
\end{gather}
Our final result for  normalized structure constant   is:
\begin{gather}
C_{\omega_1,\omega_2,\omega_3}=-i^{1/2}g^4 \frac{2}{\pi^5}\frac{\sqrt{N_c^2-1}}{N_c^2}\gamma_1^2(2+\gamma_1)^2\sqrt{\frac{G(\nu_1^*)}{\aleph'(\nu_1^*)}\frac{G(\nu_2^*)}{\aleph'(\nu_2^*)}\frac{G(\nu_3^*)}{\aleph'(\nu_3^*)}}
\Psi(\nu_1^*,\nu_2^*,\nu_3^*)\label{FinalResult}
\end{gather}
Precising the dependence on parameters \(\{\frac{g^2}{\omega_i}\}\), \(g^2\) and \(N_c\) we can write: \(C_{\omega_1,\omega_2,\omega_3}=g \frac{\sqrt{N_c^2-1}}{N_c^2} f(\frac{g^2}{\omega_1},\frac{g^2}{\omega_2},\frac{g^2}{\omega_3})\),
where \(f\) is a function which depends only on  the ratios \(\{\frac{g^2}{\omega_i}\}\). In the limit \(\gamma_i\rightarrow 0\) we get the  asymptotics (see appendix \ref{OandL} for details):
\begin{gather}
\Omega(h_1^*,h_2^*,h_3^*)\rightarrow -\frac{16 \pi^3}{\gamma_1^2\gamma_2^2\gamma_3^2}
\cdot[\gamma_1^2(\gamma_2
+\gamma_3)
+\gamma_2^2(\gamma_1+\gamma_3)+\notag\\+\gamma_3^2(\gamma_1+\gamma_2)
+\gamma_1 \gamma_2 \gamma_3)(1+O(\gamma_i))\notag\\
\Lambda(h_1^*,h_2^*,h_3^*)\rightarrow \frac{8\pi^2 (\gamma_1+\gamma_2+\gamma_3)}{\gamma_1 \gamma_2 \gamma_3}(1+O(\gamma_i))
\end{gather}
In this limit \(\gamma_i=-\frac{8g^2}{\omega_i}+o(\frac{g^2}{\omega_i})\) and the main contribution to 3-point correlator  \eqref{FinalResult} comes from the planar \(\mathcal{O}(g^2)\) term
\begin{gather}
C_{\omega_1,\omega_2,\omega_3}=-ig^2 \frac{\sqrt{N_c^2-1}}{\sqrt{2\pi}N_c^2} \frac{1}{\omega_1^\frac{5}{2}\omega_2^\frac{1}{2}\omega_3^\frac{1}{2}}(\omega_1^2(\omega_2+\omega_3)+\notag\\
\omega_2^2(\omega_1+\omega_3)+\omega_3^2(\omega_1+\omega_2)+\omega_1 \omega_2 \omega_3)(1+O(g^2)) \label{Cpert2}
\end{gather}
\begin{figure}
  \centering
  \includegraphics[scale=1]{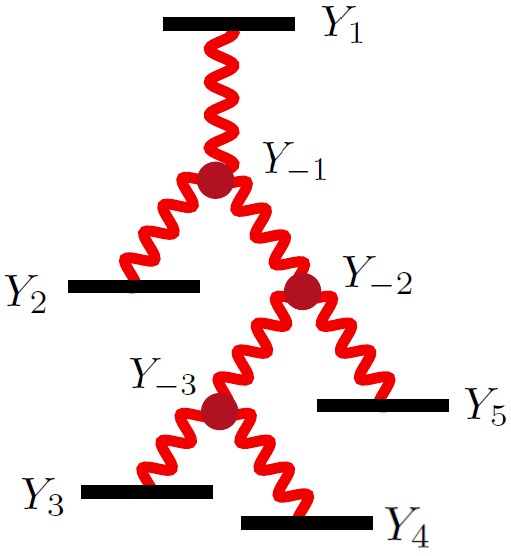}
  \caption{Typical fan diagramm which appears in calculation of correlation function. Bold black lines represent color dipoles which comes from OPE of Wilson frames and wave bold red lines - pomerons. }
  \label{fig:Nplus1}
\end{figure}
whereas the nonplanar one is \(\mathcal{O}(g^6)\). It might seem strange that the planar contribution does not start from  \(\mathcal{O}(g^4)\) terms given by the leading Feynman graphs, e.g.  with 4 gluon vertices. However, in BFKL approximation we   should keep \footnote{We thank S. Caron-Huot for the discussion on this subject.} \(\frac{g^2}{\omega}\gg\omega\). In addition, when making the point-splitting regularization we have to keep \(\frac{g^2}{\omega}|\ln (x_{31\bot}/(x-y))^2|\gg 1\). The limit \(|x_{13\bot}|\) has to be taken first, which makes the value \(g^2=0\) exceptional.  This order of limits leads to \({\cal\ O}(g^2)\) behavior of \eqref{Cpert2}.

\section{Discussion}
Our result eq.(\ref{FinalResult})  is a rare example of computation of a non-BPS structure constant. The applied BFKL approximation provides contributions from all orders in coupling constant, including infinitely many "wrapping" corrections.  Moreover, our result is valid at any \(N_c\). Since in the LO BFKL the contributions of all  fields but gluons  in \(\mathcal{N}=4 \) SYM disappear from both the definition of operators and  internal loops, the result is applicable to pure YM theory at any \(N_c\), including \(N_c=3\).  It would be interesting to apply our structure constants to the OPE at hard scattering in real QCD and to work out the full "dictionary" relating them to the OPE in the Lipatov's  2-dimensional \(SL(2,C)\) CFT --the basis of our BFKL computation.   It is also not hopeless, though challenging, to compute these structure constants in the NLO approximation in SYM. Our present result may serve as an important, all-wrappings test for the future computations of similar quantities in the integrability approach to planar AdS\(_5\)/CFT\(_4\).

One can also apply the logic of OPE expansion over Wilson lines to the more general case of 1+k -correlator where one of operaotors is oriented along \(n_+\) direction and \(k\) others - along \(n_-\). Indeed \(k\) operators along \(n_-\) form a pancake layer for the one along \(n_+\) and evolution again can be described by evolution equation. In the leading BFKL order only \(1\rightarrow 2\) vertexes contribute. So we have a cascade (see Fig.\ref{fig:Nplus1}) of BK equations which means that we should sum up over all fan type diagrams built up of \(k-1\) BK vertexes.

\begin{acknowledgments}
\label{sec:acknowledgments}
We thank J.~Bartels, S.~Caron-Huot, L.~Lipatov, J.~Penedones and V.~Schomerus for useful comments and discussions. Our special thanks  to G.~Korchemsky who participated in the initial stage of this work.  The work of E.S. and V.K. was supported by the People Programme (Marie Curie Actions) of the European Union's Seventh Framework Programme FP7/2007-2013/ under REA Grant Agreement No 317089 (GATIS).
The work of V.K.  has received funding from the European Research Council (Programme
”Ideas” ERC-2012-AdG 320769 ”AdS-CFT-solvable”), from the ANR grant StrongInt (BLANC- SIMI-
4-2011) and from the ESF grant HOLOGRAV-09- RNP- 092.
The work of I.B.  was supported by DOE contract
 DE-AC05-06OR23177 under which JSA operates Jefferson Lab and by the grant DE-FG02-97ER41028.
E.S. and V.K.
are very grateful to the Institute for Advanced Study (Princeton), where a part of this work was done,
for hospitality. V.K. also thanks the Ambrose Monell Foundation, for the generous support during his stay in Princeton.
\end{acknowledgments}

\appendix

\section{Conformal structure of correlator of 3 light ray operators }\label{ApConfStrOf3LR}

Three-point correlation function of operators with spin can be represented as a sum over different tensor structures \cite{Costa:2011mg}:
\begin{align}
<\mathcal{O}^{l_1}_{n_1}(x_1) \mathcal{O}^{l_2}_{n_2}(y_1)\mathcal{O}^{l_3}_{n_3}(z_1)>
=\sum_{m_{12}, m_{13}, m_{23}} \lambda_{m_{12},m_{23},m_{13}}
 \left[\begin{array}{ccc}\Delta_1 & \Delta_2 & \Delta_3 \\ l_1 & l_2 & l_3 \\ m_{23} & m_{13} & m_{12}\end{array}\right]
\end{align}
where $\Delta_i$ is dimension and $l_i$ is Lorentz spin. The sum runs
over positive integer numbers \(m_{ij}\) satisfying the set of inequalities:
\begin{align}
m_1=l_1-m_{12}-m_{13}\ge 0\,,\quad l_2=k_2-m_{12}-m_{23}\ge 0\,,\quad l_3=k_3-m_{13}-m_{23}\ge 0
\end{align}
In general case the tensor structure has the following form:
\begin{gather}
\left[
\begin{matrix}
\Delta_1 & \Delta_2 & \Delta_3\\
l_1 & l_2 & l_3\\
m_{23} & m_{13} & m_{12}
\end{matrix}
\right]=\notag\\
=\frac{(V_{1,23})^{l_1-m_{12}-m_{13}}(V_{2,31})^{l_2-m_{12}-m_{23}}(V_{3,12})^{l_3-m_{13}-m_{23}}(H_{12})^{m_{12}}(H_{13})^{m_{13}}(H_{23})^{m_{23}}}{P_{12}^{\frac{1}{2}(\Delta_1+\Delta_2-\Delta_3+l_1+l_2-l_3)}P_{13}^{\frac{1}{2}(\Delta_1+\Delta_3-\Delta_2+l_1+l_3-l_2)}P_{23}^{\frac{1}{2}(\Delta_2+\Delta_3-\Delta_1+l_2+l_3-l_1)}}
\end{gather}
We are interested in tensor structures integrated along three different light-ray directions with positions of local operators are restricted to one-dimensional orthogonal space. Let's choose three different light-ray direction \(n_1,\ n_2,\ n_3\) and parameterize the coordinates of local operators in the following way:
\begin{gather*}
x_1=v_1n_1+x_{1\bot},\\
x_2=v_2n_2+x_{2\bot},\\
x_3=v_3n_3+x_{3\bot}
\end{gather*}
Then
\begin{gather*}
P_{12}=-2v_1v_2<n_1,n_2> -\  x_{12\bot}^2+i\epsilon,\\
P_{13}=-2v_1v_3<n_1n_3> -\ x_{13\bot}^2+i\epsilon,\\
P_{23}=-2v_2v_3<n_2n_3> -\ x_{23\bot}^2+i\epsilon,
\end{gather*}
\begin{gather*}
H_{12}=-x_{12\bot}^2<n_1 n_2>, \ \ \ H_{13}=-x_{13\bot}^2<n_1 n_3>,\ \ \ H_{23}=-x_{23\bot}^2<n_2 n_3>,
\end{gather*}
\begin{gather*}
V_{1,23}=\frac{v_2<n_1 n_2>P_{13}-v_3<n_1 n_3>P_{12}}{P_{23}},\\
V_{2,31}=\frac{v_3<n_2 n_3>P_{12}-v_1<n_1 n_2>}{P_{13}},\\
V_{3,12}=\frac{v_1<n_1 n_3>P_{23}-v_2<n_2 n_3>P_{13}}{P_{12}}
\end{gather*}
Then integral over \(v_1,v_2,v_3\) gives us:
\begin{gather}
\int dv_1\int dv_2\int dv_3 \left[
\begin{matrix}
\Delta_1 & \Delta_2 & \Delta_3\\
l_1 & l_2 & l_3\\
m_{23} & m_{13} & m_{12}
\end{matrix}
\right]=(-1)^{m_{12}+m_{13}+m_{23}}(x_{12\bot}^2)^{m_{12}}(x_{13\bot}^2)^{m_{13}}(x_{23\bot}^2)^{m_{23}}\cdot\notag\\
\cdot\sum\limits_{q_1=0}^{l_1-m_{12}-m_{13}}\sum\limits_{q_2=0}^{l_2-m_{12}-m_{23}}\sum\limits_{q_3=0}^{l_3-m_{13}-m_{23}}(-1)^{q_1+q_2+q_3}
\times C_{l_1-m_{12}-m_{13}}^{q_1}C_{l_2-m_{12}-m_{23}}^{q_2}C_{l_3-m_{13}-m_{23}}^{q_3}\cdot\notag\\
\cdot<n_1 n_2>^{l_1-m_{13}-q_1+q_2}<n_1 n_3>^{l_3-m_{23}-q_3+q_1}<n_2 n_3>^{l_2-m_{12}-q_2+q_3}\cdot\notag\\ \cdot MI(\frac{\Delta_2+\Delta_3-\Delta_1+l_1+l_2-l_3}{2}-m_{12}+m_{23}-q_2+q_3, \frac{\Delta_1+\Delta_3-\Delta_2-l_1+l_2+l_3}{2}-\notag\\
-m_{23}+m_{13}+q_1-q_3,
\frac{\Delta_1+\Delta_2-
\Delta_3+l_1-l_2+l_3}{2}-m_{13}+m_{12}+q_2-q_1,\notag\\
 l_3-m_{13}-m_{23}-q_3+q_2,l_1-m_{12}-m_{13}-q_1+q_3,
l_2-m_{12}-m_{23}-q_2+q_1)\label{TS3dif}
\end{gather}
where we have introduced Master Integral:
\begin{gather}
MI(\alpha,\beta,\gamma,k_1,k_2,k_3)=
\int\limits_{-\infty}^{\infty}\int\limits_{-\infty}^{\infty}\int\limits_{-\infty}^{\infty}dv_1 dv_2 dv_3 \notag\\
\frac{v_1^{k_1}v_{2}^{k_2}v_3^{k_3}}{(-2v_2v_3<n_2n_3>-x_{23\bot}^2+i\epsilon)^\alpha(-2v_1v_3<n_1n_3>-x_{13\bot}^2+i\epsilon)^\beta(-2v_1v_2<n_1n_2>-x_{12\bot}^2+i\epsilon)^\gamma}
\end{gather}
which can be explicitly calculated:
\begin{gather}
MI(\alpha,\beta,\gamma,k_1,k_2,k_3)
=\frac{\pi i (-1)^{\frac{1}{2}(1+k_1+k_2+k_3)-\alpha-\beta-\gamma}}{2^{\frac{1}{2}(3+k_1+k_2+k_3)}(x_{12\bot}^2)^\gamma(x_{13\bot}^2)^\beta(x_{23\bot}^2)^\alpha}\notag \\ \left(\frac{x_{13\bot}^2}{<n_1n_3>}\right)^{\frac{1}{2}(1-k_2+k_1+k_3)}
\left(\frac{x_{12\bot}^2}{<n_1n_2>}\right)^{\frac{1}{2}(1+k_1+k_2-k_3)}\left(\frac{x_{23}^2}{<n_2n_3>}\right)^{\frac{1}{2}(1+k_2+k_3-k_1)}\notag\\
\left[(-1)^{k_1}\frac{\Gamma(\frac{1}{2}(1+k_3+k_1-k_2))}{\Gamma(\alpha)\Gamma(\beta)}\Gamma(\frac{1}{2}(-1-k_3-k_1+k_2)+\beta)
\Gamma(\alpha+\frac{1}{2}(k_1-k_3-k_2-1))\right.\cdot\notag\\ \cdot\sum\limits_{j_2=0}^{k_2}C_{k_2}^{j_2}(-1)^{j_2}\frac{\Gamma(\gamma+j_2+\frac{1}{2}(k_3-k_1-k_2-1))}{\Gamma(\gamma-k_2+j_2)}+
(-1)^{k_3}\frac{\Gamma(\frac{1}{2}(1-k_2+k_1+k_3))}{\Gamma(\beta)\Gamma(\gamma)}\cdot\notag\\
\cdot\Gamma(\frac{1}{2}(-1+k_2-k_3-k_1)+\beta)\Gamma(\gamma+\frac{1}{2}(k_3-k_1-k_2-1))\cdot\notag\\
\cdot\left.\sum\limits_{j_2=0}^{k_2}C_{k_2}^{j_2}(-1)^{j_2}\frac{\Gamma(\alpha+j_2+
\frac{1}{2}(k_1-k_2-k_3-1))}{\Gamma(\alpha-k_2+j_2)} \right] \label{MI}
\end{gather}
Now we should plug (\ref{MI}) into (\ref{TS3dif}). Turns out that all tensor structures collapse in one:
\begin{gather}
\int\int\int\left[
\begin{matrix}
\Delta_1 & \Delta_2 & \Delta_3\\
l_1 & l_2 & l_3\\
m_{23} & m_{13} & m_{12}
\end{matrix}
\right]
=C_{m_{12},m_{13},m_{23}}(\Delta_1,\Delta_2,\Delta_3,l_1,l_2,l_3)\cdot\notag\\
\cdot\frac{<n_1n_2>^{\frac{l_1+l_2-l_3-1}{2}}<n_1n_3>^{\frac{l_1+l_3-l_2-1}{2}}<n_2n_3>^{\frac{l_2+l_3-l_1-1}{2}}}
{(x_{12\bot}^2)^{\frac{1}{2}(\Delta_1+\Delta_2-\Delta_3-1)}(x_{13\bot}^2)^{\frac{1}{2}(\Delta_1+\Delta_3-\Delta_2-1)}(x_{23\bot}^2)^{\frac{1}{2}(\Delta_2+\Delta_3-\Delta_1-1)}}
\end{gather}
and it means that we can write integrated 3-point correlator as one term:
\begin{gather}
\int\int\int d v_1 d v_2 d v_3 <\mathcal{O}^{l_1}_{n_1}(x_1) \mathcal{O}^{l_2}_{n_2}(y_1)\mathcal{O}^{l_3}_{n_3}(z_1)>=\notag\\
=C(\{\Delta_i\},\{l_i\})\frac{<n_1n_2>^{\frac{l_1+l_2-l_3-1}{2}}<n_1n_3>^{\frac{l_1+l_3-l_2-1}{2}}<n_2n_3>^{\frac{l_2+l_3-l_1-1}{2}}}
{(x_{12\bot}^2)^{\frac{1}{2}(\Delta_1+\Delta_2-\Delta_3-1)}(x_{13\bot}^2)^{\frac{1}{2}(\Delta_1+\Delta_3-\Delta_2-1)}(x_{23\bot}^2)^{\frac{1}{2}(\Delta_2+\Delta_3-\Delta_1-1)}}
\end{gather}
characterizing by only one structure constant:
\begin{gather}
C(\{\Delta_i(l_i)\},\{l_i\})=\notag\\
=\sum_{m_{12}, m_{13}, m_{23}} \lambda_{m_{12},m_{23},m_{13}}(g_{_{YM}}^2,N,\{\Delta_i\},\{l_i\})C_{m_{12},m_{23},m_{13}}(\{\Delta_i\},\{l_i\})
\end{gather}

\section{Around converging points}\label{AroundConvPoints}

During the computation we took a limit of convergent points \(|x_{13}|\rightarrow 0\) as the following replacement: \(\frac{|x_1-x_3||\beta-\gamma|}{|x_1-\beta||x_3-\gamma|}\rightarrow \frac{|x_1-x_3||\beta-\gamma|}{|x-\beta||x-\gamma|}\). In this appendix we will explain the correctness of such approximation.

We have to show that the replacement as in (\ref{replacement}) leads to the difference of order \(o(|x_{13}|^{\gamma_1}|y_{13}|^{\gamma_2}|z_{13}|^{\gamma_3})\)\footnote{We remind that \(\gamma=O(\frac{g^2}{\omega})\) and \(\gamma<0\)} which can be omitted. Let's start with planar contribution and introduce a length-scale of our 3-point correlator: \(R_0\sim|x-y|_\bot\)\(\sim|y-z|_\bot\sim|z-x|_\bot\) as well as an intermediate scale \(R\) such that \(\lim\limits_{|x_{13}|\rightarrow0}\frac{|x_{13}|}{R(|x_{13}|)}=0\), \(\lim\limits_{|x_{13}|\rightarrow0}\frac{R(|x_{13}|)}{R_0}=0\). If we cut out three circles\footnote{It means that \(\alpha, \beta, \gamma\) are out of these circles.} of radiuses \(R(|x_{13}|)\), \(R(|y_{13}|)\), \(R(|z_{13}|)\) around points \(x_{1,3}\), \(y_{1,3}\), \(z_{1,3}\) correspondingly the difference between exact and approximate expressions will be of order \(o(|x_{13}|^{\gamma_1}|y_{13}|^{\gamma_2}|z_{13}|^{\gamma_3})\). To see it is enough to notice that the difference in the last brackets of the difference:
\begin{gather}
D=\left(\frac{|x_1-x_3|^2|\beta-\gamma|^2}{|x_1-\beta|^2|x_3-\gamma|^2}\right)^{\frac{1}{2}+i\nu_1}{}_2F_1 {}_2F_1-\left(\frac{|x_1-x_3|^2|\beta-\gamma|^2}{|x-\beta|^2|x-\gamma|^2}\right)^{\frac{1}{2}+i\nu_1}=\notag\\
=\left(\frac{|x_1-x_3|^2|\beta-\gamma|^2}{|x-\beta|^2|x-\gamma|^2}\right)^{\frac{1}{2}+i\nu_1}
\left(\left(\frac{|x-\beta|^2|x-\gamma|^2}{|x_1-\beta|^2|x_3-\gamma|^2}\right)^{\frac{1}{2}+i\nu_1}{}_2F_1 {}_2F_1-1\right)\label{difference}
\end{gather}
has regular power expansion:
\begin{gather}
\left(\left(\frac{|x-\beta|^2|x-\gamma|^2}{|x_1-\beta|^2|x_3-\gamma|^2}\right)^{\frac{1}{2}+i\nu_1}{}_2F_1 {}_2F_1-1\right)\rightarrow \frac{|x_{13}|}{R(|x_{13}|)}(...)+\frac{|x_{13}|^2}{R(|x_{13}|)^2}(...)+...
\end{gather}
what leads to the contribution of the form\footnote{partial derivatives come from \(D_\bot\) in (\ref{Ramki3pConfRepPlanar})}:
\begin{gather}
\partial_{x_{1}}\cdot\partial_{x_{3}}\left(\frac{|x_{13}|^{3+\gamma_1}}{R(|x_{13}|)}(...)+(\frac{|x_{13}|^{4+\gamma_1}}{R(|x_{13}|)^2})(...)+...\right)
\end{gather}
Choosing for example \(R(|x_{13}|)\sim \sqrt{|x_{13}|R_0}\) we get:
\begin{gather}
\partial_{x_{1}}\partial_{x_{3}}\left(\frac{|x_{13}|^{3+\gamma_1}}{R(|x_{13}|)}(...)+(\frac{|x_{13}|^{4+\gamma_1}}{R(|x_{13}|)^2})(...)+...\right)=o(|x_{13}|^{\gamma_1}).
\end{gather}

Thus we should just check that the contribution from configurations when at least one of \(\alpha, \beta, \gamma\) - points inside of a circle is of order  \(o(|x_{13}|^{\gamma_1}|y_{13}|^{\gamma_2}|z_{13}|^{\gamma_3})\). Let's take for example points \(x_1, x_3\) and circle of radius \(R(|x_{13}|)\) around them.

There are several configurations of \(\alpha, \beta, \gamma\). Let's first analyse the case when \(|x-\beta|<R(|x_{13}|),\ \ |x-\gamma|>R(|x_{13}|),\ \ |x-\alpha|>R(|x_{13}|) \). Let's introduce a new rescaled variable \(\sigma=\frac{|x-\beta|}{|x_{13}|}\) and estimate the contribution from circle of radius \(R(|x_{13}|)\) around \(x_{1,3}\).
In our configuration : \(\frac{|\beta-\gamma|^2}{|x-\gamma|^2}=O(1)\) and it means that after rescaling small length \(|x_{13}|\) disappears from anharmonic ratios (and thus from (\ref{Lipatov4p})) and the asymptotic form of contribution at \(|x_{13}|\rightarrow 0\) is the following:
\begin{gather}
\sim\partial_{x_1}\partial_{x_3}|x_{13}|^2\int \limits^{R(|x_{13}|)/x_{13}}\frac{\sigma d\sigma}{\sigma^{2+\gamma}} \sim\partial_{x_1}\partial_{x_3}|x_{13}|^2 \frac{(\frac{R}{|x_{13}|})^{-\gamma_1}}{\gamma_1}.
\end{gather}

We see that the main contribution comes from large distances (we also remember that \(\gamma_1<0\)) and for \(R(|x_{13}|)\sim \sqrt{|x_{13}|R_0}\) we get:
\begin{gather}
\partial_{x_1}\partial_{x_3}|x_{13}|^2 \frac{(\frac{R}{|x_{13}|})^{-\gamma_1}}{\gamma_1}=o(|x_{13}|^{\gamma_1})
\end{gather}

The next configuration: \(|x-\beta|<R(|x_{13}|),\ \ |x-\gamma|<R(|x_{13}|),\ \ |x-\alpha|>R(|x_{13}|)\). In this case we can introduce two variables \(\sigma_1=\frac{|x-\beta|}{|x_{13}|}\), \(\sigma_2=\frac{|x-\gamma|}{|x_{13}|}\) and again estimate the contribution as:
\begin{gather}
\sim\partial_{x_1} \partial_{x_3} |x_{13}|^2\frac{(\frac{R}{|x_{13}|})^{-\gamma_1}}{\gamma_1} =o(|x_{13}|^{\gamma_1})
\end{gather}

The last case, when all three points are in the one circle - \(|x-\beta|<R(|x_{13}|),\ \ |x-\gamma|<R(|x_{13}|),\ \ |x-\alpha|<R(|x_{13}|)\) gives us:
\begin{gather}
\sim \partial_{x_1} \partial_{x_3} |x_{13}|^{4+\gamma_2+\gamma_3} \left(\frac{R(|x_{13}|)}{|x_{13}|}\right)^{2+\gamma_2+\gamma_3}=o(|x_{13}|^{\gamma_1})
\end{gather}

The similar analysis can be carried out in nonplanar case. Alternatively one can notice that \(\Upsilon_{npl}\) and \(\Upsilon_{pl}\) differ just by coordinate-independent factor: \(\Upsilon_{npl}(\nu_1,\nu_2,\nu_3;x_0,y_0,z_0)=\frac{\Omega(\omega_1,\omega_2,\omega_3)}{\Lambda(\omega_1,\omega_2,\omega_3)}\Upsilon_{pl} (\nu_1,\nu_2,\nu_3;x_0,y_0,z_0)\) what means that we can rewrite nonplanar integral as a planar which was already analyzed.
\section{$\Omega$ and $\Lambda$}\label{OandL}
Planar integral:
\begin{gather}
\Upsilon_{pl} (\nu_1,\nu_2,\nu_3;x_0,y_0,z_0)=\notag\\
=\int\int\int\frac{d^2\alpha d^2\beta d^2\gamma}{|\gamma-\beta|^2 |\gamma-\alpha|^2|\beta-\alpha|^2}
\frac{(|\beta-\gamma|^2)^{\frac{1}{2}+i\nu_1}}{(|x_0-\beta|^2)^{\frac{1}{2}+i\nu_1}(|x_0-\gamma|^2)^{\frac{1}{2}+i\nu_1}}\cdot\notag\\
\cdot\frac{(|\alpha-\gamma|^2)^{\frac{1}{2}+i\nu_2}}{(|\alpha-y_0|^2)^{\frac{1}{2}+i\nu_2}(|\gamma-y_0|^2)^{\frac{1}{2}+i\nu_2}}
\frac{(|\alpha-\beta|^2)^{\frac{1}{2}+i\nu_3}}{(|\alpha-z_0|^2)^{\frac{1}{2}+i\nu_3}(|\beta-z_0|^2)^{\frac{1}{2}+i\nu_3}}\label{UpsilonPlInt}
\end{gather}
Nonplanar:
\begin{gather}
\Upsilon_{npl} (\nu_1,\nu_2,\nu_3;x_0,y_0,z_0)=\notag\\
=\int\int\frac{d^2\beta d^2\gamma}{|\gamma-\beta|^4}
\frac{(|\beta-\gamma|^2)^{\frac{1}{2}+i\nu_1}}{(|x_0-\beta|^2)^{\frac{1}{2}+i\nu_1}(|x_0-\gamma|^2)^{\frac{1}{2}+i\nu_1}}\cdot\notag\\
\cdot \frac{(|\beta-\gamma|^2)^{\frac{1}{2}+i\nu_2}}{(|y_0-\beta|^2)^{\frac{1}{2}+i\nu_2}(|y_0-\gamma|^2)^{\frac{1}{2}+i\nu_2}}
\frac{(|\beta-\gamma|^2)^{\frac{1}{2}+i\nu_3}}{(|z_0-\beta|^2)^{\frac{1}{2}+i\nu_3}(|z_0-\gamma|^2)^{\frac{1}{2}+i\nu_3}}
\end{gather}
They have the same coordinate dependence:
\begin{gather}
\Upsilon_{pl} (\nu_1,\nu_2,\nu_3;x_0,y_0,z_0)=\frac{\Omega(\nu_1,\nu_2,\nu_3)}{|x_0-y_0|^{\Delta_{12}}|x_0-z_0|^{\Delta_{13}}|y_0-z_0|^{\Delta_{23}}},\notag\\
\Upsilon_{npl} (\nu_1,\nu_2,\nu_3;x_0,y_0,z_0)=\frac{\Lambda(\nu_1,\nu_2,\nu_3)}{|x_0-y_0|^{\Delta_{12}}|x_0-z_0|^{\Delta_{13}}|y_0-z_0|^{\Delta_{23}}}
\end{gather}
and the planar can be rewritten as nonplanar:
\begin{gather}
\Upsilon_{pl} (\nu_1,\nu_2,\nu_3;x_0,y_0,z_0)=\frac{\Omega(\nu_1,\nu_2,\nu_3)}{\Lambda(\nu_1,\nu_2,\nu_3)}\Upsilon_2 (\nu_1,\nu_2,\nu_3;x_0,y_0,z_0)\label{PlanarCherezNonplanar}
\end{gather}
Explicit expression for \(\Omega(h_1,h_2,h_3)\) and \(\Lambda(h_1,h_2,h_3)\)  was obtained in \cite{Korchemsky:1997fy}
 (here we present formulas in relevant for us \(n=0\) case: \(h_i=\frac{1}{2}+i\nu_i\)):
\begin{gather}
\Omega(h_1,h_2,h_3)=\pi^3 \left(\Gamma^2(h_1)\Gamma^2(h_2)\Gamma(1-h_1)\Gamma(1-h_2)\Gamma(1-h_3) \right)^{-1}\times\notag\\
\times\sum\limits_{a=1}^3J_{a}(h_1,h_2,h_3)\bar{J}_a (h_1,h_2,h_3),\label{OmegaExplicit}
\end{gather}
where
\begin{gather}
J_1=\Gamma(h_1+h_2-h_3)\frac{\Gamma(h_1)\Gamma(1-h_1)}{\Gamma(h_2)\Gamma(1-h_2)}G_{44}^{24}\left(1\left|\begin{matrix} &h_2, 1-h_2,h_3,h_3\\ &0,0,-h_1+h_3,-1+h_1+h_3\end{matrix} \right. \right),\notag\\
\bar{J_1}=\frac{\Gamma(1-h_1)\Gamma(1-h_2)}{\Gamma(1-h_3)\Gamma(1-h_1-h_2+h_3)}G_{44}^{33}\left(1\left|\begin{matrix} &h_2, 1-h_2,h_3,h_3\\ &0,0,-h_1+h_3,-1+h_1+h_3\end{matrix} \right. \right)\notag\\
J_2=\frac{\Gamma(h_1+h_2-h_3)\Gamma(1-h_1)\Gamma(h_1)\Gamma(1-h_2)\Gamma(h_2)\Gamma^2(1-h_3)}{\Gamma(1+h_1-h_3)\Gamma(2-h_1-h_3)}\times\notag\\
\times {}_4F_3\left( \left.\begin{matrix} &h_2, 1-h_2,1-h_3,1-h_3\\ &1,2-h_1-h_3,1+h_1-h_3\end{matrix} \right| 1 \right),\notag
\end{gather}
\begin{gather}
\bar{J_2}=\frac{\Gamma(1-h_1)\Gamma(h_3)}{\Gamma(h_2)\Gamma(1-h_1-h_2+h_3)}G_{44}^{42}\left(1\left|\begin{matrix} &h_2, 1-h_2,h_3,h_3\\ &0,0,-h_1+h_3,-1+h_1+h_3\end{matrix} \right. \right),\notag\\
J_3(h_1,h_2,h_3)=J_2(h_2,h_1,h_3), \ \ \ \ \ \ \ \ \ \ \ \bar{J}_3(h_1,h_2,h_3)=\bar{J}_2(h_2,h_3,h_3) \notag
\end{gather}
and
\begin{gather}
\Lambda(h_1, h_2, h_3)=\pi^2 2^{2(h_1+h_2+h_3)-4}\frac{\Gamma(1-h_1)}{\Gamma(h_1)}\frac{\Gamma(1-h_2)}{\Gamma(h_2)}\frac{\Gamma(1-h_3)}{\Gamma(h_3)}\cdot\notag\\
\cdot \frac{\Gamma(-\frac{1}{2}+\frac{1}{2}(h_1+h_2+h_3))}{\Gamma(\frac{3}{2}-\frac{1}{2}(h_1+h_2+h_3))}
\frac{\Gamma(\frac{1}{2}(h_1+h_2-h_3))}{\Gamma(1-\frac{1}{2}(h_1+h_2-h_3))} \cdot\notag\\
\cdot\frac{\Gamma(\frac{1}{2}(h_1-h_2+h_3))}{\Gamma(1-\frac{1}{2}(h_1-h_2+h_3))} \frac{\Gamma(\frac{1}{2}(-h_1+h_2+h_3))}{\Gamma(1-\frac{1}{2}(-h_1+h_2+h_3))}
\end{gather}
At \(\gamma_i\rightarrow 0\) asymptotics of \(\Lambda\) can be easily calculated:
\begin{gather}
\Lambda(1+\frac{\gamma_1}{2}, 1+\frac{\gamma_2}{2}, 1+\frac{\gamma_3}{2})\rightarrow \frac{8\pi^2 (\gamma_1+\gamma_2+\gamma_3)}{\gamma_1 \gamma_2 \gamma_3}(1+O(\gamma_i))
\end{gather}
Asymptotic form  of \(\Omega\) at small \(\gamma_i\) can be extracted from explicit expression (\ref{OmegaExplicit}):
\begin{gather}
\Omega(1+\frac{\gamma_1}{2}, 1+\frac{\gamma_2}{2}, 1+\frac{\gamma_3}{2})\rightarrow -\frac{16 \pi^3}{\gamma_1^2\gamma_2^2\gamma_3^2}
\cdot[\gamma_1^2(\gamma_2
+\gamma_3)
+\gamma_2^2(\gamma_1+\gamma_3)+\notag\\+\gamma_3^2(\gamma_1+\gamma_2)
+\gamma_1 \gamma_2 \gamma_3)(1+o(\gamma_i))
\label{Omega}
\end{gather}
Or alternatively it can be obtained directly from the integral representation (\ref{UpsilonPlInt}):
\begin{eqnarray}
&&\hspace{-1mm}
{\Omega(1-\epsilon_1,1-\epsilon_2,1-\epsilon_3)
\over  (x^2)^{1+\epsilon_1-\epsilon_2-\epsilon_3}}
\nonumber\\
&&\hspace{-1mm}
=~
\int\! {d^2w_1 d^2w_2d^2w_3\over (w_{12}^2)^{\epsilon_3}(w_{13}^2)^{\epsilon_2}(w_{23}^2)^{\epsilon_1}}
\Big({1\over (w_1-x)^2(w_2-x)^2}\Big)^{1-\epsilon_3}
\Big({1\over w_1^2w_3^2}\Big)^{1-\epsilon_2}
~=~
\label{fla1}
\end{eqnarray}
which is obtained from Eq.  (\ref{UpsilonPlInt}) by taking $z_0=0$ and performing the inversion.
It is convenient to multiply Eq. (\ref{fla1}) by $x_i$ and split the integral in two parts
\begin{eqnarray}
&&\hspace{-11mm}
x_i{\Omega(1-\epsilon_1,1-\epsilon_2,1-\epsilon_3)
\over  (x^2)^{1+\epsilon_1-\epsilon_2-\epsilon_3}}
~=~I_i(x,\epsilon_1,\epsilon_2,\epsilon_3)+I_i(x,\epsilon_1,\epsilon_3,\epsilon_2)
\end{eqnarray}
where
\begin{eqnarray}
&&\hspace{-1mm}
I_i(x,\epsilon_1,\epsilon_2,\epsilon_3)
\nonumber\\
&&\hspace{-1mm}
=~\!\int\! {d^2w_1 d^2w_2d^2w_3\over (w_{12}^2)^{\epsilon_3}(w_{13}^2)^{\epsilon_2}(w_{23}^2)^{\epsilon_1}}
(x-w_1)_i
\Big({1\over (w_1-x)^2(w_2-x)^2}\Big)^{1-\epsilon_3}
\Big({1\over w_1^2w_3^2}\Big)^{1-\epsilon_2}
\label{fla3}
\end{eqnarray}
It is clear that at $\omega_i=0$ the integral over $w_1$ diverges only as $w_1\rightarrow 0$ and integral over $w_2$ is either UV divergent as $w_2\rightarrow x$ or IR as $w_2\rightarrow\infty$.
In any case, divergence comes from the region $w_2\gg w_1$ so at first we will calculate the 3-point CF
\begin{eqnarray}
&&\hspace{-11mm}
\!\int\! {d^2w_3\over (w_{13}^2)^{\epsilon_2}(w_{23}^2)^{\epsilon_1}(w_3^2)^{1-\epsilon_2}}
\label{int1}
\end{eqnarray}
at $w_2\gg w_1$, then perform integral over $w_2$ and finally over $w_1$.

The explicit calculation of the integral (\ref{int1}) yields at $w_2\gg w_1$:
\begin{eqnarray}
&&\hspace{-1mm}
\!\int\! {d^2w_3\over (w_{13}^2)^{\epsilon_2}(w_{23}^2)^{\epsilon_1}(w_3^2)^{1-\epsilon_2}}
~=~{\pi\over (w_2^2)^{\epsilon_1}}\Big[\ln{w_2^2\over w_1^2}+{1\over\epsilon_1}+{1\over\epsilon_2}
+2\psi(1)
\label{int1otvet}\\
&&\hspace{-1mm}
-~\psi(1+\epsilon_1)-\psi(1-\epsilon_1)+2\psi(1)-\psi(1+\epsilon_2)-\psi(1-\epsilon_2)
+O\big({(w_1,w_2)\over w_2^2}\big)R(\epsilon_1)R(\epsilon_2)\Big]
\nonumber
\end{eqnarray}
where $R(\epsilon_i)$ is a non-singular function of $\epsilon_i$.

Using integral (\ref{int1otvet})  we get
\begin{eqnarray}
&&\hspace{-5mm}
I_i(x,\epsilon_1,\epsilon_2,\epsilon_3)~=~\!\int\! d^2w_1(x-w_1)_i
\Big({1\over (w_1-x)^2}\Big)^{1-\epsilon_3}
\Big({1\over w_1^2}\Big)^{1-\epsilon_2}
\\
&&\hspace{-1mm}
\times~\!\int\! d^2w_2 {1\over (w_{12}^2)^{\epsilon_3}[(x-w_2)^2]^{1-\epsilon_3}(w_2^2)^{\epsilon_1}}
~\Big[\ln{w_2^2\over w_1^2}+{1\over\epsilon_1}+{1\over\epsilon_2}+4\psi(1)
\nonumber\\
&&\hspace{3mm}
-~\psi(1+\epsilon_1)-\psi(1-\epsilon_1)-\psi(1+\epsilon_2)-\psi(1-\epsilon_2)
+O\big({(w_1,w_2)\over w_2^2}\big)R(\epsilon_1)R(\epsilon_2)\Big]
\nonumber
\end{eqnarray}
The integration over $w_2$ has the form
\begin{eqnarray}
&&\hspace{-11mm}
\int\! d^2w_2{\ln w_2^2a\over [(x-w_2)^2]^{1-\epsilon}[(w_{12})^2]^{\epsilon_3} (w_2^2)^{\epsilon_1}}
~=~~{\pi\Gamma(\epsilon_3)\Gamma(\epsilon_1)\over (x^2)^{\epsilon_1}\Gamma(\epsilon_1+\epsilon_3)}
{\Gamma(1-\epsilon_3)\Gamma(1-\epsilon_1-\epsilon_3)
\over\Gamma(1-\epsilon_1)}
\nonumber\\
&&\hspace{-11mm}
\times~
\big[\ln x^2a+{1\over\epsilon_1}-{1\over\epsilon_1+\epsilon_3}-\psi(1+\epsilon_1)-\psi(1-\epsilon_1)+\psi(1+\epsilon_1+\epsilon_3)+\psi(1-\epsilon_1-\epsilon_3)\big]
\label{C15}
\end{eqnarray}
where $a$ is an arbitrary constant. Performing the remaining trivial integral over $w_1$ one obtains
\begin{eqnarray}
&&\hspace{-1mm}
{\pi^3x_i\over (x^2)^{1+\epsilon_1-\epsilon_2-\epsilon_3}}
{\Gamma(\epsilon_1)\Gamma(\epsilon_2)\Gamma(\epsilon_3)\over \Gamma(\epsilon_1+\epsilon_3)}
\Big[{2\over\epsilon_1}+{2\over\epsilon_2}-{1\over\epsilon_1+\epsilon_3}
+4\psi(1)-2\psi(1+\epsilon_1)-2\psi(1-\epsilon_1)
\nonumber\\
&&\hspace{11mm}
+~4\psi(1)-\psi(1+\epsilon_2)-\psi(1-\epsilon_2)
-~2\psi(1)
+\psi(1+\epsilon_1+\epsilon_3)+\psi(1-\epsilon_1-\epsilon_3)-2\psi(1)
\nonumber\\
&&\hspace{22mm}
+~\psi(1+\epsilon_2+\epsilon_3)+\psi(1-\epsilon_2-\epsilon_3)\Big]
{\Gamma(1+\epsilon_3)\Gamma(1-\epsilon_2-\epsilon_3)\Gamma(1-\epsilon_1-\epsilon_3)\over \Gamma(1-\epsilon_1)\Gamma(1-\epsilon_2)\Gamma(1+\epsilon_2+\epsilon_3)}
\nonumber
\end{eqnarray}
At small $\epsilon_i\ll 1$ this turns to
\begin{equation}
\hspace{-1mm}
I_i(x,\omega_1,\omega_2,\omega_3)~
=~{x_i\over (x^2)^{1+\epsilon_1-\epsilon_2-\epsilon_3}}{\pi^3\over \epsilon_1\epsilon_2\epsilon_3}
\Big[{2(\epsilon_1+\epsilon_3)\over\epsilon_1}+{2(\epsilon_1+\epsilon_3)\over\epsilon_2}-1\Big]
\big[1+O(\epsilon_i)\big]
\label{C16}
\end{equation}
and therefore
\begin{equation}
\hspace{-1mm}
\Omega(1-\epsilon_1,1-\epsilon_2,1-\epsilon_3)
~=~{2\pi^3\over \epsilon_1\epsilon_2\epsilon_3}\Big[1+{\epsilon_1+\epsilon_2\over\epsilon_3}+{\epsilon_2+\epsilon_3\over\epsilon_1}+{\epsilon_1+\epsilon_3\over\epsilon_2}\Big]\big[1+O(\epsilon_i)\big]
\label{C17}
\end{equation}
which coincides with Eq. (\ref{Omega}) after identification \(\epsilon_i=-\frac{\gamma_i}{2}\).

\bibliographystyle{JHEPb}



\printindex

\end{document}